\pdfoutput=1
\documentclass[twocolumn,5p]{elsarticle}
\usepackage{color}
\usepackage{listings}
\usepackage{graphicx} 
\usepackage{multirow}
\usepackage[table,xcdraw,dvipsnames]{xcolor}
\usepackage{titlesec}
\usepackage{paralist}
\usepackage{siunitx}
\usepackage{rotating}
\usepackage{eqparbox}
\usepackage{wrapfig}
\usepackage{mathtools}
\usepackage{xspace}
\usepackage{enumitem}
\usepackage{hyperref}

\usepackage[framemethod=tikz]{mdframed}
\usepackage{lipsum}
\usetikzlibrary{shadows}
\newmdenv[tikzsetting= {fill=white!20},roundcorner=10pt, shadow=true]{myshadowbox}
\usepackage{graphics}
\usepackage{colortbl} 
\usepackage{multirow} 
\usepackage{balance}
\usepackage{picture}
\usepackage{soul}
\usepackage{fourier} 
\usepackage{array}
\usepackage{makecell}

\usepackage{times}
\usepackage{wasysym}

\usepackage{verbatim}
\usepackage{algorithm}
\usepackage{algorithmicx}
\usepackage{algpseudocode}
\usepackage[export]{adjustbox}
\renewcommand{\footnotesize}{\scriptsize}
\definecolor{lightgray}{gray}{0.8}
\definecolor{darkgray}{gray}{0.6}

\usepackage[table]{xcolor}
\definecolor{Gray}{rgb}{0.88,1,1}
\definecolor{Gray}{gray}{0.85}
\definecolor{Blue}{RGB}{0,29,193}

\usepackage{pgfplots}
\usepackage{pgfplotstable}
\pgfplotsset{compat=newest}
\usepackage{caption}

\newcommand{\rvt}[1]{
}

\definecolor{MyDarkBlue}{rgb}{0,0.08,0.45} 
\newcommand{\quart}[4]{\begin{picture}(100,6)
{\color{black}\put(#3,3){\circle*{4}}\put(#1,3){\line(1,0){#2}}}\end{picture}}
\newcommand{\quartr}[4]{\begin{picture}(100,6)
{\color{black}\put(#3,3){\color{red}\circle*{6}}\put(#1,3){\line(1,0){#2}}}\end{picture}}

\newcommand{\quartex}[4]{
\begin{picture}(25,6)
    {
        \color{black}
        \put(#3,3)
        {\circle*{4}}
        \put(#1,3)
        {\line(1,0){#2}}
    }
\end{picture}
}

\definecolor{Gray}{gray}{0.95}
\definecolor{LightGray}{gray}{0.975}
\newcommand{\bi}{\begin{itemize}}
\newcommand{\ei}{\end{itemize}}
\newcommand{\be}{\begin{enumerate}}
\newcommand{\ee}{\end{enumerate}}

\newcommand{\fig}[1]{Figure~\ref{fig:#1}}
\newcommand{\tab}[1]{Table ~\ref{tab:#1}}

\newcommand{\sway}{{SWAY\xspace}}
\newcommand{\rv}[1]{{\color{black}{#1}}}

\usepackage{url}

\begin{document}

	\begin{frontmatter}
		
		\title{Beyond Evolutionary Algorithms for Search-based Software Engineering}
		
		\author{Jianfeng Chen}
		\ead{jchen37@ncsu.edu}
		\author{Vivek Nair\corref{cor1}}
		\ead{vivekaxl@gmail.com}
		\author{Tim Menzies}
		\ead{tim.menzies@gmail.com}
		
		\cortext[cor1]{Corresponding author:
		email: vivekaxl@gmail.com
		Tel:+1-919-523-920(Vivek)}
		\address{Department of Computer Science, North 
		Carolina State University, Raleigh, NC, USA}
		\pagenumbering{arabic}

            \begin{abstract} 
            
            \noindent
			{\bf Context:}
           Evolutionary algorithms typically require large number of evaluations (of solutions) to converge -- which can be very slow and expensive to evaluate.
			
			\noindent 
			{\bf Objective:} 
			To solve search-based software engineering (SE) problems, using fewer evaluations than evolutionary methods.
			
			\noindent
			{\bf Method:} 
            Instead of mutating a small population, we build a very large initial population which is then culled using a recursive bi-clustering  chop approach.
            We evaluate this approach on multiple SE models, unconstrained as well as constrained, and compare its performance with standard evolutionary algorithms.
			
			\noindent
			{\bf Results:}  
            Using just a few evaluations (under 100), we can obtain  comparable results to state-of-the-art evolutionary algorithms.
			
			\noindent
			{\bf Conclusion:}
            Just because something works, and is widespread use, does not necessarily mean that there is no value in seeking methods to improve that method. Before undertaking search-based SE optimization tasks using traditional EAs, it is recommended to try other techniques, like those explored here, to obtain the same results with fewer evaluations.

\end{abstract}
		
	\end{frontmatter}
	\pagenumbering{arabic} 

\section{Introduction}\label{sect:intro}
Due to the complexities of software architectures and shareholder requirements, it is often hard to 
solve complex modeling problems via
a standard numerical mathematical analysis or some deterministic algorithms~\cite{Harman12}.
There are many reasons for this complexity:
\bi
\item When procedural code is used within the model of a domain,
every ``if'' statement can divide the internal problem space into different regions (once for each branch in the ``if''). Such software models
cannot be optimized via traditional numerical methods which assume  models
are a single continuous differentiable function.
\item
Finding solutions to problems  often means  accommodating competing choices. When stakeholders
propose multiple   goals, search-based SE (SBSE) methods
can reflect on goal interactions to
propose novel solutions to hard optimization problems
such as configuring products in complex
product lines~\cite{sayyad13b}, tuning parameters of a data miner~\cite{tantithamthavorn2016automated},
or finding best configurations
for clone detection algorithms~\cite{Wang:Harman13}.
\ei
For these tasks,  many SBSE researchers usually use
evolutionary
algorithms (EA)~\cite{sayyad13b,tantithamthavorn2016automated,Wang:Harman13}. Evolutionary algorithms start by generating a set of initial solutions and improve them through  crossover and mutation, also known as reproduction operators. They are inspired by  evolution in  nature and make no parametric
assumptions about  problems being generated. In our experience, 
this has made them particularly well-suited for SE problems.
However, evolutionary algorithms typically require large number of evaluations (of solutions) to converge.
Real-world model-based applications may be very expensive to evaluate (with respect to computation time, resources required etc.).

So, can we do better than EA for SBSE? Or, are there faster alternatives to EA?
This paper experimentally evaluates one such  alternative called SWAY (short for
the \underline{S}ampling \underline{WAY}): 
\be
\item Similar to a standard EA, {\em generate} an initial population;  
\item  
Intelligently {\em select} a cluster within the population generated with best scores.
\ee
\sway{} runs so fast since it terminates  
after just
$O(lg N)$ evaluations of $N$ candidate solutions.
 \sway{}'s  intelligent selection mechanism for exploring subsets of the 
population is a recursive binary chop  that  (i)~finds and evaluates only the two most
dissimilar examples, then (ii)~recurses only on  half of the data containing the better among its similar example.
As shown later in this paper, for this process to work, it is important to have the right
definition of  ``dissimilar''.

Note the differences between \sway{} and standard EA:
\begin{enumerate}
\item
SWAY quits after the initial generation while EA reasons over multiple
generations;
\item
SWAY  makes no use of reproduction operators so
there is no way for lessons learned to accumulate
as it executes;
\item Depending on the algorithm, not all members
of the population will be evaluated -- e.g. active
learners~\cite{krall2015gale}  only
evaluate a few representative individuals.
\end{enumerate}
Because of the limited nature of this search, until recently,
we would have dismissed SWAY as comparatively less effective than EA
for exploring multi-goal optimization.
Nevertheless, quite by accident, we have stumbled onto evidence that
has dramatically changed our opinion about SWAY. Recently
we were working  
with an algorithm called GALE~\cite{krall2015gale}.
 GALE  is an evolutionary algorithm that includes SWAY as a sub-routine:
 \[
 \begin{array}{rcl}
 \textit{evolution}& = &\textit{generations} * 
   \left\{
    \begin{array}{l}
    \textit{mutation} \\
    \textit{crossover} \\
    \textit{sampling}\\    
    \end{array}\right.\\

 \textit{SWAY}& = &\textit{GALE} - \textit{evolution}\\
 \end{array}
 \]
 While porting GALE from Python to Java, we accidentally 
  disabled evolution. 
To our surprise,  the  ``broken'' version  
 worked   as well, or better, than the original GALE. This is an interesting result since
GALE has been compared against dozens of models in a recent TSE article~\cite{krall2015gale}  and dozens more in Krall's Ph.D. thesis~\cite{krall2014faster}.
In those studies, GALE was found to be competitive against widely used evolutionary algorithms.
If Krall's work is  combined with the results from our accident, then we conjecture
that   the success of GALE is due less to   ``evolution''
  than to   ``sampling''   many options. 
  This, in turn, could lead to a new generation of very
  fast optimizers since, as we show below, sampling can be much faster than evolving.
  
 The rest of this paper describes SWAY and presents evidence for its utility.
While we have tested
SWAY on the standard EA  benchmarks such as 
DTLZ, Fonseca, Golinski, Srinivas, etc.~\cite{Deb2001}, those
results are not
included here since, in our experience, results from those benchmarks
are less convincing to the SE community than results from software models.
Hence, here we present results from:
 \bi
 \item POM3: a model of agile teams selecting their next task from the scrum backlog~\cite{port08,1204376};
 \item XOMO: a  model predicting software development time, effort, risks and defects~\cite{me07f,me09a,me09e};
 \item MONRP: a model of next release planning that recommends which functionality to code next~\cite{bagnall2001next}.
 \ei
After presenting some background motivational notes, this paper offers general
notes on multi-objective evolutionary algorithms. This is followed by a description of SWAY and the POM3, XOMO, MONRP models. Experimental results are then presented
showing that SWAY achieves results competitive with standard methods
(NSGA-II and SPEA2)  using orders of magnitude fewer evaluations.
Working with the MONRP models, we also find that a seemingly minor detail
(the implementation of the distance function used to recognize 
``dissimilar'' examples) is of vital importance to the success of SWAY.
Finally, this paper concludes with     experiments on ``super-charging'' that tests whether SWAY
can boost the performance of standard optimizers.  

Our observations after conducting the study are:
\bi
\item The mutation strategies seen in a recently published EA algorithm~(GALE) adds little value;
\item GALE without evolution~(SWAY) runs an order of magnitude faster than EAs;
\item Optimization found by SWAY are similar to those found by SBSE algorithms;
\item How  we   recognize  
``dissimilar'' examples is of vital importance;
\item Super-charging (combining SWAY with standard SBSE optimizers) is not
useful.
\ei
More generally, our
 conclusion is that
{\em sampling is an interesting research approach for multi-dimensional
optimization that deserves further attention by the SBSE community.}

\subsection{Connection to Prior Work}

This paper significantly extends prior work by the authors. The background
notes in the next section are new to the paper, as is the super-charging
study. Also, this paper
repairs a significant drawback seen in  initial   describing of SWAY.
At SSBSE'16~\cite{nair2016accidental}, we demonstrated how \sway{} can be used to find near optimal solutions for problems like XOMO and POM. While
an interesting result, it turns out that the early definition of ``dissimilar'' used
by the earlier version of \sway{}  was  only applicable to problems whose decision space is constrained in nature. The results on other types of problems, were less than impressive. In this paper, we expose the weakness of the  earlier variant of \sway{} and show other definitions of ``dissimilar'' can make \sway{} very useful for
other domains.

\subsection{When is SWAY most Useful, Useless?}\label{sect:useful}
\rv{
SWAY is designed as a fast substitution of EAs for solving SBSE problems.
It can avoid large amount of model evaluations, which are very common in previous evolutionary algorithms. In view of this, SWAY is particularly useful in  following two scenarios.
}

\sway{} would be most useful
if it is proposed
to put humans-in-the-loop to help guide the evaluations (e.g. as done in~\cite{reed2007using}). In this scenario, standard EAs
might have to ask a human for up to $O(N^2)$ opinions for $G$ generations.
On the other hand,
\sway{}  would only trouble the user $O(lg N)$ times
 
Also,
\sway{} was created to solve problems, where the practitioner is not able to  evaluate  thousands of individuals for e.g. Wang et al.~\cite{wang2013searching} spent 15 years of CPU time to find software clone detectors or model explored by Krall et al.~\cite{krall2015gale}) which take hours to perform a single evaluation. 

\rv{
However, as discussed later in this paper,  SWAY has two core assumptions.
Firstly, it is applicable only when there is a mapping 
between ``genotype'' and ``phenotype'' space;
i.e. between the settings to the model inputs and outputs of the model. 
Even though such mapping may not exist in every model, we find here that for  SE models (that
were written with the explicit goal of effecting
outputs with input decisions), this
assumption
holds adequately, at least
for the purposes of improving model output. 

Secondly,  SWAY techniques for dividing the data makes the {\em spectral learning assumption};
i.e. that within the raw dimensions of data seen in any domain, there exists a small set
of {\em spectral} dimensions which can usefully approximate the larger set~\cite{Kamvar03}.
While the universality of the spectral assumption has not been proven,
it has seen to hold in many domains; e.g. see any data analysis method that uses
principle components analysis~\cite{arias2017spectral,bellet2016spectral,shi2017spacecraft,redgate2016principal,li2016randomized}.
}

\subsection{Access to Code}
For the purposes of reproducibility, all the code
and data used in this paper are 
available at http://tiny.cc/Sway.

\section{Frequently Asked Questions}

Before exploring the technical details on SWAY, we digress to  answer
some frequently asked questions about this research.

\subsection{What is the Value of Seeking Simplicity?}

While SWAY does not necessarily produce better optimizations,  we advocate its
use since it is  very simple and very fast. 
But {\em what is the value of reporting simple and faster ways to achieve results that are currently achievable by slower and more complex methods?}

In terms of core science, we argue that the better can we understand something, the better we can match tools to SE. Tools which are poorly matched to task are usually complex and/or slow to execute.  SWAY seems a better match for the tasks explored in this paper since it is neither complex nor slow. Hence, we argue that SWAY is interesting
in terms of its core scientific contribution to SE optimization research.

Seeking simpler and/or faster solutions is not just theoretically interesting.
It is also an approach currently in vogue in contemporary software engineering.
Calero and Pattini~\cite{calero2015green} comment that   ``redesign for greater simplicity'' also motivates much contemporary industrial work.  In their survey of modern SE companies, they find that many current  organizational  redesigns are motivated (at least in part) by arguments based on ``sustainability'' (i.e. using fewer resources to achieve results). According to  Calero and Pattini, sustainability is now a new source of innovation. Managers used sustainability-based redesigns to explore cost-cutting opportunities. In fact, they say, sustainability is now viewed by many companies as a mechanism for gaining complete advantage over their competitors.  Hence, a manager might
ask a programmer to  assess methods like SWAY as a  technique to generate new
and more interesting products.

\subsection{Why not just use more of the Cloud?}
SWAY reduces the number of evaluations required to optimize a model (and hence the CPU cost of this kind of analysis by one to two orders of magnitude). Given the ready availability of cloud-based CPU, we are sometimes asked 
{\em about the benefits of merely making something run 100 times faster
when we can just buy more CPU on the cloud?} 

In reply, we say that CPUs are not an unlimited resource that should be
applied unquestionably to computationally expensive problems.
\bi
\item
We can no longer rely on Moore's
Law~\cite{658762} to double our computational power
very 18 months.
Power consumption and heat
dissipation issues effectively block further
exponential increases to CPU clock
frequencies~\cite{kuman03}.  
\item
Even if we could build
those faster CPUs, we would still need to power them.
CPU power requirements (and the
pollution associated with generating that
power~\cite{thib14}) is now a significant issue.
Data centers consume 1.5\% of globally electrical
output and this value is predicted to grow
dramatically in the very near
future~\cite{koomey08}  (data centers in the
USA used 91 billion kilowatt-hours of electrical
energy in 2013, and they will be using 139 billion
kilowatt-hours by 2020 (a 53\%
increase)~\cite{thib14}).
\item
Even if (a)~we could
build faster CPUs and even if (b)~we had the energy to
power them and even if (c)~we could dispose of the
pollution associated with generating that energy,
then all that CPU+energy+pollution offset would be a service
that must be paid for. Fisher et al.~\cite{Fisher:2012}
comment that cloud
computation is a heavily monetized environment that charges for all
their services (storage, uploads, downloads, and CPU time).
While each small part of that service is cheap, the total annual
cost to an organization can be exorbitant.
Google reports that a 1\%
reduction in CPU requirements saves them millions of dollars in power costs.
\ei
Hence we say that tools like SWAY, which use less CPU, are interesting because they
let us achieve the same goals with fewer resources.

\section{Multi-objective Evolutionary Algorithms (MOEA)}
SWAY is a multi-objective optimization algorithm. This section offers a general
background to the general area of MOEAs.

SBSE involves utilizing the rich literature of search based optimization to solve software engineering problems. The automatic evaluation of solutions opens a whole range of possibilities with EAs. In last few years, EAs have been used to solve problems in software engineering like requirement selection~\cite{durillo2011study}, resource allocation in project scheduling~\cite{luna2014software} etc. 

EAs are very flexible and can be used for variety of problems. There are only two ingredients for adapting EAs to SE problems:
\begin{enumerate}
    \item{The choice of representation of the problem}
    \item{The definition of the fitness function}
\end{enumerate}
This simplicity and ready applicability has led to wide adoption of the EAs in the SE domain.

The SE problem can be conceptualized as an optimization problem where the function to be optimized is unknown. 
A general multi-objective algorithm can be posed as follows:

\begin{equation}\label{eq:opti}
    \min_{x} F(x) = [F_1(x), F_2(x), ... , F_k(x)]^T
\end{equation}
where $x$ is a vector in decision space and $k$ is the number of objectives. 

In contrast to single-objective (where $k=1$ in equation~\ref{eq:opti}), in many times, there is no single global solution to a multi-objective problem but rather a set of points that all fit a predetermined definition of an optimum. The predominant concept in defining optimal point is called Pareto optimality, which is defined as:
\begin{quote}
$x$ {\it dominates} $y$ iff for every objective, $x$ performs no worse than $y$ and there exist some objective(s), $x$ outperforms $y$.
\end{quote}

This is also known as binary domination. 
Also, Zitzler et al.~\cite{Zitzler04indicator-basedselection} proposed indicator-based domination, which rather than binary domination does not return $\{\mathit{True, False}\}$ but rather return a measure of dominance of a solution over other.

A standard EA can be described as follows (similar to one described in section 1):
\begin{itemize}
    \item{Generate initial population of solutions using a initialization policy, such as random strategy}
    \item{Evaluate the solution using the problem specific fitness function}
    \item{Repeat till pre-defined stopping criterion is true:
    \begin{itemize}
        \item{Create new population using problem specific reproduction operators}
        \item{Evaluate the newly generated population}
        \item{Select solutions from the new population for the next generation. The selection  mimics {\it ``survival of 
        the fittest''}.}
    \end{itemize}
    }
\end{itemize}

Depending on the selection strategy, most MOEAs can be classified into:
\begin{itemize}
    \item{\textbf{Pareto Dominance Based:}}
    Pareto dominance based algorithms uses the binary domination to select solutions for the successive generations. These techniques are used in tandem with niching operators to preserve the diversity among the solutions. \\
    Examples: NSGA-II~\cite{Deb00afast}, PAES~\cite{knowles1999pareto}, SPEA2~\cite{Zitzler01spea2:improving}
    
    \item{\textbf{Decomposition Based:}}
    Decomposition based algorithms divide the problem into a set of sub-problems, which are solved simultaneously in a collaborative manner. Each sub-problem is usually an aggregation of the objectives with uniformly distributed weight vectors.\\
    Examples: MOGLS~\cite{jaszkiewicz2002performance}, MOEA/D~\cite{zhang2007moea}
    
    \item{\textbf{Indicator Based:}}
    Indicator based methods work by establishing a complete order among the solutions using a single scalar metric like hypervolume etc.\\
    Examples: HypE~\cite{bader2011hype}, IBEA~\cite{zitzler2004indicator}
\end{itemize}

All above algorithms typically evaluate thousands to millions of individuals as part of their execution. A fundamental challenge in engineering and other domains is that evaluation of a solution is very expensive:
\begin{itemize}
    \item
        Zuluaga et al. comment on the cost of evaluating all decisions for their models of software/hardware co-design:
        ``synthesis of only one design can take hours or even days.'' ~\cite{Zuluaga:13}.
    \item
        Krall et al. explored the optimization of complex NASA models of air traffic control. After discussing the simulation needs of NASA's research scientists,
        they concluded that  those models would take three months to execute, even utilizing NASA's supercomputers~\cite{krall15:hms}.
\end{itemize}

The intention of  SWAY is  to  reduce  that  running  time and cost  without  sacrificing  the  quality  of results. 

\begin{figure}[!t]
\small
\hspace{0.4cm}\begin{lstlisting}[mathescape,frame=r,numbers=right]
func SWAY(candidates):
  if |candidates| < $\epsilon$
    return candidates
  else
    [west,westItems], [east,eastItems] $\leftarrow$ split(candidates)
    if west $\succeq$ east
        return SWAY(westItems)
    if east $\succeq$ west
        return SWAY(eastItmes)
    if $\neg$(west $\succeq$ east $\lor$ east $\succeq$ west)
        return SWAY(westItems) $\cup$ SWAY(eastItems)
\end{lstlisting}
\caption{Framework of \sway{} algorithm. {\color{blue}\bf SWAY} recursively splits  candidates until
number of candidates is small enough (In this paper, we let threshold $\epsilon$ be $\sqrt{N}$).
The $\succeq$ in Line 6, 8 or 10 indicates {\it dominates}, or {\it better than}.
See \S\ref{sect:principle} for details of {\color{blue}\bf split}.
}
\label{fig:sway_frame}  
\end{figure}

\section{SWAY}\label{sec:sway}
This section describes SWAY, the multi-objective optimization algorithm.
After this, the rest of the paper conducts experimental evaluations of SWAY.

\subsection{Overview}\label{sect:swayo}
\sway{}, short for the \underline{S}ampling \underline{WAY}, is designed as an optimizer for SBSE problems.
Unlike  common EAs, which search for  optimal points by evolution,  \sway{} first randomly generates large number
of candidates, recursively divides the candidates and only selects ones.
Figure~\ref{fig:sway_frame} describes  framework of \sway{} algorithm. In Line 5, the input candidates is divided into two parts--{\it westItems} and {\it eastItems}; each of them contains  representative(s).
Next, such representatives are compared to each other: if the east part representative dominates west part representative, then all candidates in {\it westItems} are discarded; vice versa (Line 6-11).

Obviously, the time complexity as well as effectiveness of \sway{} highly dependent on {\it split} function. In \sway{}, candidates are clustered based on
decision space; that is, in \sway{}, we cluster the promising/unpromising candidates through their decisions.
Model evaluations are only applied to very limited representative(s) of each cluster -- for purpose of
deciding which cluster contains promising candidates.
Consequently, as  a recursive process, \sway{} only requires O(lg N) model evaluations, where N is number of initial candidates.

However, designing {\it split} is a challenging task of using \sway{}. An inappropriate {\it split} function can lead to
invalid clustering of candidates-- cannot distinguish between promising and unpromising candidates. This is one of the limitation of \sway{}. In next section, we will introduce some principles of designing
{\it split} as well as two  strategies used in our experiment.

Another limitation of \sway{} is that the clustering process is based on decision space, or genotype space. If, for some specific
model, there exist no mapping between genotype and phenotype space, then finding a {\it split} function connecting
genotype space and phenotype space is not feasible.
``Mersenne twister function''~\cite{matsumoto1998mersenne} is an extreme example. In case one 
model is some combination of random functions, then  \sway{}  will fail. However, our experience and experiments showed that these random models are very rare in SBSE.


\subsection{Principles and guides to split function}\label{sect:principle}

This section introduces some principles and guidelines for customizing the 
{\it split} function from  Figure~\ref{fig:sway_frame} to different domains.

From \S\ref{sect:swayo}, we note that the purpose of {\it split} is to divide candidates into two parts; a good part and a second poorer part, which  is expected to be discarded.
To demystify this splitting process,
we list several principles of {\it split} function as follows:

\begin{enumerate}[label=(\Roman*)]
\item Split must run over the information available {\em before} candidate evaluation; i.e.
split must run over the decision space
instead of objective space;
\item After  {\em split} similar candidates should be clustered together.
\item Further, dissimilar, or opposite  candidates should be separated into different clusters.
\item For each subspace of candidates' decision space, candidates are expected to be separated into two clusters, instated of gathering in one cluster.
\end{enumerate}

Principle (I) is for overcoming the computing-complexity of model evaluations in SBSE problems.
Principles (II) and (III) assume: there are some mappings between genotype and phenotype space (described in \S\ref{sect:intro}) in
SBSE problems-- candidates with similar decisions should have similar objectives. This assumption lets us avoid evaluating every candidate.
Principle (IV) is to guarantee the diversity of \sway{} outputs.
For problems with multi-objectives, diversity of results is a main consideration.

To help researchers understand these principles intuitively, here we have a guide
for creating  {\it split} functions:
\begin{enumerate}
\item To support principle~II and III, for problems with continuous decisions or discrete decisions (with interval attribute~\cite{stevens1946theory}), Minkowski distance, especially the Euclidean distance, is a good choice for measuring the similarity between candidates.
\item If clustering all candidates into two groups fails, then it is very likely that
it did not fulfill principle (IV). To overcome this, we can exploit domain knowledge for the problem,
manually divide candidates into several groups first. After that, apply the {\it split} function to each group.
\item Main goal of {\it split} function is to divide candidates into two clusters. To determine which cluster is good one, we must select representative(s). Two possible strategies for this selection --
first is selecting random points; another is selecting extreme points, which can form the diameter of decision space.
\end{enumerate}

In \S\ref{sect:split1} - \ref{sect:split2}, we show how to apply these guides in real instances.

\subsection{Split function for continuous models (POM, XOMO)}
\label{sect:split1}
This section shows how to customize a {\it split} function for POM and XOMO problems,
which are used later in our experiments (for details see \S\ref{sect:models}).
For POM and XOMO, the decision space is continuous.
Following guide 2, we can apply Euclidean distance to measure similarity among two candidates.
After defining the similarity measure, we also need to find a way to gather/separate candidates. We suggest researchers to first try FASTMAP~\cite{faloutsos1995fastmap} .
The FASTMAP heuristic randomly picks a solution  and then find  extreme points within the decision space. All solutions are then projected onto the line joining the extreme points. The solutions are split into two groups based on their projection on the line. 
 
Much
prior work has shown that FASTMAP is an effective  and fast separator
~\cite{krall14aaai,krall15:hms, nair2016accidental}.
It makes  {\em spectral learning assumption}; i.e.  the raw dimensions
of a data set can be better characterized by a smaller number of underlying {\em spectral}
dimensions. We note that any analysis that uses a principle component analysis makes such an spectral assumption~\cite{bellet2016spectral,shi2017spacecraft,redgate2016principal}. 

\begin{figure}[!t]
\small
\hspace{0.4cm}\begin{lstlisting}[mathescape,frame=r,numbers=right]
func FASTMAP(candidates): 
  rand $\leftarrow$ random selected candidate
  E $\leftarrow$ furthest(rand, candidates) 
  W $\leftarrow$ furthest(E, candidates)     
  for X $\in$ candidates
    X$_d \leftarrow \overrightarrow{\text{WE}}\cdot\overrightarrow{\text{WX}}/||\overrightarrow{\text{WE}}||$  # map X to the line W~E
  WestItems $\leftarrow \emptyset$
  EastItems $\leftarrow \emptyset$
  for X $\in$ candidates
    if X$_d < \frac{1}{2}||\overrightarrow{\text{WE}}||$
        WestItems $\leftarrow$ WestItems $\cup$ X
    else
        EastItems $\leftarrow$ EastItems $\cup$ X
  return W, E, WestItems, EastItems 
\end{lstlisting}
\caption{Pseudocode of FASTMAP, the splitting strategy in POM and XOMO models in this paper. FastMap~\cite{faloutsos1995fastmap} maps all candidates into one line which is connecting two extreme points. See \S\ref{sect:split1} for details.
}
\label{fig:cont_split}  
\end{figure}

Figure~\ref{fig:cont_split} is the pseudocode of FASTMAP algorithm.
It first finds out two extreme points (line 2-4), and then map every other candidates into the line connecting above extreme points (line 5-6).
That is, we reduce decision space into one dimension (X$_d$ in line 6). Analogous to PCA algorithm,
this line can be treated as the first principle component of the decision space.
In line 9-13, candidates are clustered according to X$_d$.
Finally, following guide 3, we select  extreme points as the representatives for each cluster.

\subsection{Split function for discrete models (MONRP)}\label{sect:split2}
MONRP problem (for details see \S\ref{sect:models}) has a discrete decision space. In our previous attempts, simply applying algorithm in Figure~\ref{fig:cont_split} did not work for this problem.
From guide 2, we can exploit domain knowledge to divide candidates into several groups first and then use algorithm in Figure~\ref{fig:cont_split} in each group.

Let us explore how to exploit domain knowledge of the problem. 
 In  MONRP,
if more features are to release in early versions, then developing teams' workload tends to be intense in early stage (otherwise, the workload can shift to late stage, or spread equally).
Exploiting  this domain knowledge,   we can divide candidates into groups according to their workload mode:
\begin{equation}
\mathbf{WL(y)} = ||\{1\leq i \leq N:y_i < \text{P}/2\}||
\end{equation}

\label{rvt:2b}\rvt{2b} where $\mathbf{y}\in [1,P]^N$ represents a release plan for a project with N features and have P releases at maximum.
Consequently, $\mathbf{WL(y)}$ represents how many features will be released in first half of  release plan. We can divide candidates into groups. 
Note that, once an initial split
of the space has completed using this principle,
we can further split the resulting space using FASTMAP 
because after manual division, each group contains a subspace of the decision space (principle (IV)) which can be more easily handled by {\it split} functions described in Figure~\ref{fig:cont_split}.

This is just one guidance for discrete models. For other problems, researchers can adjust the pre-handler according to problem description. In our experience,  these pre-processors are   simple to code, just like $\mathbf{y}$.

\section{Models}\label{sect:models}
\subsection{Unconstrained Models - Continuous}

\begin{figure*}[t]
        \centering
        \begin{minipage}{0.35\paperwidth}
        { \footnotesize
        \begin{tabular}{l|lrr|lr}
              &\multicolumn{3}{c|}{ranges}      &\multicolumn{2}{c}{values}\\\hline
        project&feature&low&high&feature&setting\\\hline
         &rely&3&5&tool&2\\
        FLIGHT:&data&2&3&sced&3\\
         &cplx&3&6&&\\
        JPL's flight&time&3&4&&\\
        software&stor&3&4&&\\
        &acap&3&5&&\\
        &apex&2&5&&\\
        &pcap&3&5&&\\
        &plex&1&4&&\\
        &ltex&1&4&&\\
        &pmat&2&3&&\\
        &KSLOC&7&418&&\\ 
        \multicolumn{6}{c}{~}\\
        \multicolumn{6}{c}{~} \\\hline
        &prec&1&2&data&3\\
        OSP:&flex&2&5&pvol&2\\
         &resl&1&3&rely&5\\
        Orbital space &team&2&3&pcap&3\\
        plane nav\&&pmat&1&4&plex&3\\
        gudiance&stor&3&5&site&3\\
          &ruse&2&4&&\\
         &docu&2&4&&\\
        &acap&2&3&&\\
        &pcon&2&3&&\\
        &apex&2&3&&\\
        &ltex&2&4&&\\
        &tool&2&3&&\\
        &sced&1&3&&\\
        &cplx&5&6\\
        &KSLOC&75&125\\
        \end{tabular}\hspace{2cm}} 
        \end{minipage}~~~~~~~~~\begin{minipage}{0.40\paperwidth}
        {\footnotesize 
        \centering
        \begin{tabular}{l|lrr|lr}
              &\multicolumn{3}{c|}{ranges}      &\multicolumn{2}{c}{values}\\\hline
        project&feature&low&high&feature&setting\\\hline
         &rely&1&4&tool&2\\
        GROUND:&data&2&3&sced&3\\
         &cplx&1&4&&\\
        JPL's ground&time&3&4&&\\
        software&stor&3&4&&\\
         &acap&3&5&&\\
        &apex&2&5&&\\
        &pcap&3&5&&\\
        &plex&1&4&&\\
        &ltex&1&4&&\\
        &pmat&2&3&&\\
        &KSLOC&11&392&&\\
        \multicolumn{6}{c}{~} \\\hline
         &prec&3&5&flex&3\\
        OSP2: &pmat&4&5&resl&4\\
         &docu&3&4&team&3\\
        OSP&ltex&2&5&time&3\\
        version 2&sced&2&4&stor&3\\
         &KSLOC&75&125&data&4\\
         &&&&pvol&3\\
         &&&&ruse&4\\
         &&&&rely&5\\
        &&&&acap&4\\
        &&&&pcap&3\\
        &&&&pcon&3\\
        &&&&apex&4\\
        &&&&plex&4\\
        &&&&tool&5\\
        &&&&cplx&4\\
        &&&&site&6
        \end{tabular}
        } 
        \end{minipage} 
        \caption{Four project-specific XOMO scenarios~\cite{turner03}.
If an attribute can be varied, then it is mutated over the range
{\em low} to {\em high}. Otherwise it is fixed to one {\em setting}.}\label{fig:xomocases}
\end{figure*}

\subsubsection{XOMO} This section summarizes XOMO. For more details, refer to ~\cite{me07f,me09a,me09e,krall2015gale}.  
XOMO combines four software process models from Boehm's group at the University of Southern California.
XOMO's inputs are the   project descriptors of \fig{xomocases} which   can 
(sometimes) be changed
by management decisions. For example, if a manager wants to 
 (a)~{\em relax schedule pressure}, they set {\em sced} to its minimal value;
(b)~to {\em reduce functionality}  they 
halve the value of {\em kloc} and reduce minimize the size of the project
database (by setting {\em data=2});
(c)~to {\em reduce quality} (in order to race something
to market) they might move to lowest reliability, minimize the documentation work
and the complexity of the code being written,
reduce the schedule pressure to some middle value. In the language
of XOMO, this last change would be {\em rely=1, docu=1, time=3, cplx=1}.

XOMO derives  four objective scores: 
(1)~project {\em risk}; 
(2)~development {\em effort}; (3)~predicted {\em defects}; (4)~total {\em months} of development ({\em Months} = {\em effort} / {\em \#workers}).
Effort and defects are predicted from mathematical models derived from 
data collected from hundreds
of commercial and  Defense Department projects~\cite{boehm00b}. 
 As to the {\em risk} model, this model contains
 rules that triggers when management decisions decrease the 
odds of successfully completing a project: e.g.  demanding
{\em more}  reliability ({\em rely}) while  {\em decreasing} analyst capability ({\em acap}).
Such a project is ``risky'' since it means the manager is demanding more reliability from less skilled analysts.
XOMO measures {\em risk} as the percent of triggered rules.

The optimization goals for XOMO are to {\em reduce} all these values. 
\bi
\item Reduce risk;  
\item Reduce effort;
\item Reduce defects;
\item Reduce months.
\ei
Note that this is a non-trivial problem since the objectives listed above are non-separable and  conflicting in nature.
For example,
{\em increasing}
software reliability   {\em reduces} the
  number of added defects while {\em increasing} the 
software development effort.
Also, {\em more} documentation can improve team communication and {\em decrease} the number of introduced defects.
However, such increased documentation {\em increases} the development effort.

\begin{figure*}[t] 
\scriptsize

  \centering
    \begin{tabular}{|l|l|p{3in}|c|}
        \hline
        Short name &Decision             & Description         &Controllable                                        \\ \hline
        Cult&Culture              & Number (\%) of requirements that change. & yes \\\hline
        Crit&Criticality           & Requirements cost effect for safety critical systems. & yes\\\hline
        Crit.Mod&Criticality Modifier & Number of (\%) teams affected by criticality.   & yes           \\ \hline
        Init. Kn&Initial Known        & Number of (\%) initially known requirements.             & no     \\ \hline
        Inter-D&Inter-Dependency     & Number of (\%) requirements that have interdependencies.  Note that dependencies are requirements within
the {\em same} tree (of requirements), but interdependencies are requirements that live in {\em different} trees.   & no            \\\hline
        Dyna&Dynamism             & Rate of how often new requirements are made. & yes                    \\ \hline
        Size&Size            & Number of base requirements in the project.& no \\        \hline
        Plan&Plan                 & Prioritization Strategy (of requirements): one of
        0= Cost Ascending;  1= Cost Descending; 2= Value Ascending; 3= Value Descending;
        4 = $\frac{\mathit{Cost}}{\mathit{Value}}$ Ascending.
 & yes \\\hline
     T.Size&Team Size            & Number of personnel in each team   & yes                         \\ 
        \hline
    \end{tabular}
    \caption {List of inputs to POM3.
    These inputs come from Turner \& Boehm's   analysis of factors
    that control how well organizers can react to agile development practices~\cite{turner03}.
The optimization task is to find settings for the controllables in the last column.
}\label{fig:pom3decisions}
\end{figure*}

\subsubsection{POM3-- A Model of Agile Development:}\label{sec:pom3pom3} 
According to Turner and Boehm~\cite{turner03},    agile managers
struggle   to  balance   {\em idle rates}, {\em completion rates} and {\em overall cost}. 
\bi
\item
In the agile world, projects terminate after achieving a {\em completion rate} of   $(X<100)$\% of its required tasks.
\item
Team members  become {\em idle} if forced to wait for a yet-to-be-finished task from other teams. 
\item
To lower {\em idle rate} and increase {\em completion rate}, management can hire staff--but this increases  {\em overall cost}.
\ei 
Hence, in this study, our optimizers tune the 
  decisions of \fig{pom3decisions} in order to
\bi
\item Increase completion rates;
\item Reduce idle rates;
\item Reduce overall cost.
\ei 
Those inputs  are used by the POM3 model to compute
completion rates, idle times and overall cost.
For full details POM3 see~\cite{port08,1204376}. For a synopsis,
see below.

To understand POM3~\cite{port08,1204376}, consider a set of intra-dependent requirements.
 A single requirement consists
    of a prioritization {\em value} and a {\em cost}, along with a list of
    child-requirements and dependencies.  Before any requirement can be
    satisfied, its children and dependencies must first be satisfied.
    POM3 builds a requirements heap with prioritization values,
    containing
     30 to 500 requirements, with costs from 1 to 100 (values 
     chosen in consultation with Richard Turner~\cite{turner03}). 
Since POM3 models agile projects, the   {\em cost, value} figures are
constantly changing (up until the point when the requirement
is completed, after which they become fixed).

Now imagine
a mountain of requirements   hiding below the surface of a lake; i.e. it
is
 mostly invisible. As the project progresses,
the lake dries up and the mountain slowly appears. 
Programmers standing on the shore study the mountain.
Programmers are organized into teams. 
Every so often, the teams pause to plan their next sprint.
At that time, the   backlog of tasks 
comprises   the visible  requirements. 

\begin{figure*}[!t]
\footnotesize
\begin{center}
    \begin{tabular}{r|p{1.2in}|p{1.2in}|p{1.2in}}
                     & POM3a                         & POM3b             &POM3c       \\ 
                             & A broad space of projects. & Highly critical small projects& Highly dynamic large projects\\\hline
        Culture              & 0.10 $\leq x \leq$ 0.90       & 0.10 $\leq x \leq$ 0.90  & 0.50 $\leq x \leq$ 0.90  \\ 
        Criticality          & 0.82 $\leq x \leq$ 1.26       & 0.82 $\leq x \leq$ 1.26   & 0.82 $\leq x \leq$ 1.26  \\ 
        Criticality Modifier & 0.02 $\leq x \leq$ 0.10       & 0.80 $\leq x \leq$ 0.95 & 0.02 $\leq x \leq$ 0.08   \\ 
        Initial Known        & 0.40 $\leq x \leq$ 0.70       & 0.40 $\leq x \leq$ 0.70  & 0.20 $\leq x \leq$ 0.50  \\ 
        Inter-Dependency     & 0.0   $\leq x \leq$ 1.0       & 0.0   $\leq x \leq$ 1.0  & 0.0   $\leq x \leq$ 50.0 \\ 
        Dynamism             & 1.0   $\leq x \leq$ 50.0      & 1.0   $\leq x \leq$ 50.0  & 40.0   $\leq x \leq$ 50.0 \\ 
        Size                 & x $\in$ [3,10,30,100,300] & x $\in$ [3, 10, 30]     & x $\in$ [30, 100, 300]   \\ 
        Team Size            & 1.0 $\leq x \leq$ 44.0        & 1.0 $\leq x \leq$ 44.0  & 20.0 $\leq x \leq$ 44.0    \\ 
        Plan                 & 0 $\leq x \leq$ 4             & 0 $\leq x \leq$ 4    & 0 $\leq x \leq$ 4       
\end{tabular}
\end{center}

\caption{Three specific POM3 scenarios. }\label{fig:POM3abcd}
\end{figure*}

For their next sprint, 
teams prioritize work  using one of five prioritization methods:  
(1)~cost ascending;  (2)~cost descending; (3)~value ascending; (4)~value descending; (5)~$\frac{\textit{cost}}{\textit{value}}$ ascending.
Note that prioritization might be sub-optimal due to the
changing nature of the requirements {\em cost, value} as the unknown
nature of the remaining requirements.
Another wild-card that POM3
has contains an {\em early cancellation probability}
that can cancel a project after $N$ sprints~(the
value directly proportional to number of sprints). Due
to this wild-card, POM3's teams are always racing to deliver as
much as possible before being re-tasked.  The final total cost is a function
    of:
    \bi
    \item[(a)] Hours worked taken from the {\em cost} of the requirements; \item[(b)] The salary of the developers: less
    experienced developers get paid less;
    \item[(c)]  Criticalness
 of  software: mission critical software costs
    more since they are allocated more resources for software quality tasks.
   \ei

\subsection{MONRP: A Discrete Constrained Model}\label{monrp}
\begin{figure*}[!t]
\footnotesize
    \centering
    \begin{tabular}{r|c|c|c|c|c|c}
    Name     & \# Requirements & \# Releases & \# Clients & \# Density & \# Budget & Level of \\
    & & & & & & Constraints\\
    \hline
    MONRP-50-4-5-0-110 & 50                                           & 4                                        & 5                                       & 0                                       & 110                                    & 1                                                 \\
    MONRP-50-4-5-0-90 & 50                                           & 4                                        & 5                                       & 0                                       & 90                                    & 2                                                 \\
    MONRP-50-4-5-4-110  & 50                                           & 4                                        & 5                                       & 4                                       & 110                                     & 3                                                 \\
    MONRP-50-4-5-4-90  & 50                                           & 4                                        & 5                                       & 4                                       & 90                                     & 4                                                
    \end{tabular}
\caption{Variants of MONRP used in this study.}\label{fig:monrp}
\end{figure*}

In requirement engineering, next release problem (NRP) is one of problems with high complexity. 
NRP concerns with defining which requirements should be implemented for the next version of the systems, according to customer satisfaction, budget constraints as well as precedence constraints between various requirements. 
Durillo et al.~\cite{durillo2009study}, treated the next release problem as a multi-objective problem, since higher customer satisfaction and less development time or cost are conflicting objectives, we call this formulation as multi-objective NRP (MONRP).  MONRP  in this paper considers (maximizing) combination of importance and risk, (minimizing) cost and (maximizing) satisfaction.

The problem can be mathematically described as follows.
Given a software project with $N$ requirements, find a vector $\vec{y}\in\{0,\ldots,P\}^N$
so that following objectives can be optimized.

\begin{equation}
  \left\{
   \begin{aligned}
f_1 &= \sum_{j=1}^{M}t_j\left(\sum_{i=1}^{N}(P+1-y_i)I_{ij}+r_i\right)\\
f_2 &= \sum_{i=\{[1,N]|y_i>0\}} c_i \\
f_3 &= \sum_{i=\{[1,N]|y_i>0\}}\sum_1^M I_{ij}
 \end{aligned}
   \right.
\end{equation}

subject to
\[
\sum_{i=1}^{N} c_iy_i \le \text{BR}_k, \forall k\in [1,P]
\]
\[
y_i \le y_j, \forall e_{ij}\in E(G)
\]

Where,  $P$ is total number of releases; $y_i=0$ indicates abortion of requirement $i$; $M$ is number of customers and $t_j$ is the importance of t-th customer for developing company; $I$ is a matrix which element $I_{ij}$ indicators the business value of requirement $i$ in view of customer $j$; $r_i$ is the risk of requirement $i$;$c_i$ is the economic cost of achieving requirement $i$; BR is an vector showing the budget of each release; G is a DAG indicating the release topology of different
requirements.

First objectives $f_1$ indicates a combination of customer values as well as risk. The objective is to fulfill  customers' requirements, as well as requirements with high importance first.
While second objective $f_2$ sums up total economic cost for all developed requirements and $f_3$ sums up customers satisfaction at the end of all releases.

The first constraint indicating that the total cost should not exceed the budget allocated to each release.
The second constraint is for topology of  requirements.

\section{Scenarios:}

Our studies execute XOMO and POM3 (unconstrained models) in the context of seven  project-specific scenarios and MONRP (constrained model) in the context of four scenarios.

For XOMO, we use
four scenarios taken from NASA's Jet Propulsion
Laboratory~\cite{me09a}.  As shown in \fig{xomocases}, FLIGHT and GROUND
is a general description of all JPL flight and ground software while 
OSP and OPS2 are two versions of the   flight guidance system
of the Orbital Space Plane.

For POM3, we explore three scenarios proposed by Boehm.   As shown in \fig{POM3abcd}: POM3a
   covers a wide range of projects;  POM3b represents
   small and highly critical projects and POM3c represent
   large projects that are highly dynamic (ones where cost
   and value can be altered over a large range).
   
For MONRP, we explore 4 variants of MONRP, ranging from the least constrained to the most constrained. Figure~\ref{fig:monrp} lists the variants of the problem used in the paper. For example, problem variant MONRP-50-4-5-4-110, describes the scenario where a software project has 50 requirements; among all requirements, 4\% are dependent  on others; also, the software is to develop for 5 clients within 110\% of budgets. This means that the project is over-funded and hence making it less constraint wrt. its budget. The column {\it Level of Constraints} represents the level of constraints, 1 being the least constrained and 4 being the most constrained.

\begin{table*}
    \small
    \centering
        \caption{Runtimes(minutes) for applying NSGA-II algorithms (with 20 generations and 2,000 model evaluations in total)}\label{tab:runtime}
    \begin{tabular}{|c|c|c|c|}
    \hline
    Scenarios     & (T1) Total runtime & (T2) Time spent in model evaluations & T2/T1\\\hline
    Pom3a & 13.8 & 13.2& 95.0\%\\\hline
    Pom3b & 3.05 & 2.83 & 92.9\%\\\hline
    Pom3c & 33.5 & 32.3 & 96.5\%\\\hline
    XOMO-FLIGHT & 3.03 & 2.53 &89.0\%\\\hline
    XOMO-GROUND & 3.21 & 3.05 & 94.8\% \\\hline
    XOMO-OSP & 2.43 & 2.2 & 90.4\% \\\hline
    XOMO-OSP2 & 3.56 & 3.31 & 93.0\% \\\hline
    MONRP-50-4-5-0-110 & 20.3 &17.8 & 87.7\%  \\\hline
    MONRP-50-4-5-0-90 & 24.2 & 21.2 & 87.6\%\\\hline
    MONRP-50-4-5-4-110 & 26.0 & 23.9 & 91.7\%\\\hline
    MONRP-50-4-5-4-90 & 20.5& 18.5& 90.2\%\\\hline
    \rowcolor{lightgray}
    Total &153 &140.8 & 92\%(avg)\\\hline
    \end{tabular}
    
\end{table*}

\subsection{Optimizers}\label{optimizers}

In this paper we consider   
 NSGA-II~\cite{Deb00afast} and SPEA2~\cite{zit02} along with GALE (for models with continuous decision space). 
 We use   NSGA-II and SPEA2  since:
 \bi
 \item
 In his survey of the SSBE literature
in the period 2004 to 2013, Sayyad~\cite{sayyad13c}
found
25 different algorithms.
Of those,  
NSGA-II~\cite{Deb00afast} or SPEA2~\cite{zit02} were used
four times as often as anything else. 
\item
This trend  can be observed in  the more recent SSBSE'16, where NSGA-II and SPEA2 were used three times as often
as any other EA~\footnote{\url{http://tiny.cc/ssbse17_survey}}.
\ei

SPEA2's~\cite{zit02} {\em selection} sub-routine  favors individuals that dominate the most number of other solutions that are not nearby (and to break ties, it favors 
items in low density regions).

 NSGA-II~\cite{Deb00afast} 
uses a non-dominating sorting procedure to divide the solutions into {\em bands} where {\em band}$_i$ dominates all of the solutions in {\em band}$_{j>i}$. NSGA-II's
elite sampling favors the
least-crowded solutions in the better bands. 

GALE~\cite{krall2015gale} only applies domination to   two distant individuals $X,Y$.
If either dominates, GALE ignores the half of the population near to the dominated individual
and recurses on the point near the non-dominated individual.

\label{rvt:2a} \rvt{2a}It is worth pointing out that it is necessary to round decimal values into integers when some variable is in normal scale, such as the {\it ``plan''} in POM models. This strategy has been applied in previous papers~\cite{lekkalapudi2014cross,galepaper}.

\rv{
Also, a frequently asked question is why not apply NSGA-II non-dominating sorting to all initial evaluated  candidates of \sway{}.
In reply, we point to 
\tab{runtime}~\footnote{Tested in a Linux machine with 1.4GHz CPU and 4GB memory}
and observe that NSGA-II is so
slow for large sets of candidates. For example, for MONRP models, it took two hours in model evaluations (not including non-dominating sorting), compared to half an hour in standard MOEA (overall process). Consequently, in real practice, we do not
recommend
applying
non-dominating sorting to candidate sets.
}

\subsection{Performance  Measures}\label{performance_m}

We use the following   three metrics to evaluate the quality of results produced by the optimizers.
\textbf{\#Evaluations}  are the number of times an optimizer
 calls a model or evaluate a model. Note that we use evaluation calls, and not
 raw CPU time, in order to be fair to all the implementations of our system.
 Some of our implementations are very new (e.g. SWAY) while others have
 been extensively profiled and optimized by multiple research teams. \tab{runtime} also shows that more than 85\% of runtime were spent in model evaluations.

Secondly performance measure used is  Deb's \textbf{Spread} 
calculator~\cite{Deb00afast} includes the term 
$\sum_i^{N-1} (d_i - \overline{d})$  where $d_i$ is the distance between
adjacent solutions and $\overline{d}$ is the mean of all such values.  A
good spread makes all the distances equal ($d_i
\approx \overline{d}$), in which case Deb's spread
measure would reduce to some minimum value.  

Thirdly, \textbf{HyperVolume}  measure was first proposed in ~\cite{zitzler1998multiobjective} to quantitatively compare the outcomes of two or more MOEAs. Hypervolume can be thought of as ``size of volume covered''. 

Note that
hypervolume and spread are computed from the population which is returned when
these optimizers terminate. Also,
 {\em higher} values of hypervolume are {\em better} while
 {\em lower} values of spread and \#evaluations are {\em better}.

These results were  
studied using non-parametric tests (non-parametric testing in SBSE was endorsed by
Arcuri and Briand at ICSE'11~\cite{arcuri11}).
For testing statistical significance,
we used non-parametric bootstrap test 95\% confidence~\cite{efron93} followed by
an A12 test to check that any observed differences were not trivially small effects;
i.e. given two lists $X$ and $Y$, count how often there are larger
numbers in the former list (and there there are ties, add a half mark):

\[a=\forall x\in X, y\in Y\frac{\#(x>y) + 0.5*\#(x=y)}{|X|*|Y|}\]
(as per Vargha~\cite{Vargha00}, we say that a ``small'' effect has $a <0.6$). 
Lastly, to generate succinct reports, we use the Scott-Knott test to recursively
divide our optimizers. This recursion used A12 and bootstrapping  
to group together subsets that are (a)~not significantly different and are (b)~not
just a small effect different to each other. This use of Scott-Knott is endorsed
by Mittas and Angelis~\cite{mittas13}
and by Hassan et al.~\cite{7194626}.

\section{Experiments}
\subsection{Research Questions}
We formulate our research questions in terms of the applicability of the techniques used in \sway{}. As our approach promotes sampling instead of evolutionary techniques, it is a natural question, ``how is this possible?''. ``Sampling instead of evolution'' is very counter-intuitive to  practitioners since, EAs have been widely accepted and adopted by the SBSE community. 

Also, if we are only sampling from 100/10000 solutions, it is possible to miss solutions, which are not present in the initial population. These are valid arguments while trying to find near-optimal solutions to problems with competing objectives. It may be argued that sampling is such a straight forward approach so it is critical to evaluate the effectiveness of such sampling method on variety of SE models both constrained and unconstrained, discrete as well as continuous. It is also interesting to see how the techniques of \sway{} can be borrowed by the traditional MOEAs to improve the performance. 

Therefore, to assess feasibility of our algorithm, we must consider:
\begin{itemize}
    \item Performance scores generated from \sway{} when compared to other MOEAs
    \item Can \sway{} be used to super-charge other MOEAs such that the performance of super-charged MOEA is better than standard MOEA?
\end{itemize}

\begin{figure*}
 \resizebox{0.975\textwidth}{!}{
\centering
    \begin{minipage}{0.5\paperwidth}
    
    {\small   \begin{tabular}{l@{~~~~}l@{~~~~}r@{~~~~}r@{~~}c@{}r}
    
      \multicolumn{1}{l}{\textbf{Rank}}& \textbf{using} & \textbf{med.} & \textbf{IQR} & \\ 
      
    \rowcolor{lightgray}\arrayrulecolor{lightgray}
    \textbf{Pom3a} & \textbf{} & \textbf{} & \textbf{} & \\\hline
      1 &        SWAY 2 &    91  &  22 & \quart{36}{10}{44}{0} \\
      1 &         GALE &    99  &  18  & \quart{48}{9}{48}{0} \\
      1 &        SWAY 4 &    104  &  13  & \quart{46}{6}{51}{0} \\
    \hline  2 &       NSGA-II &    151   &  13 & \quart{73}{6}{74}{0} \\
      2 &        SPEA 2 &    156  &  17 & \quart{71}{8}{76}{0} \\
    \hline  
    \rowcolor{lightgray}\arrayrulecolor{lightgray}
    \textbf{Pom3b} & \textbf{} & \textbf{} & \textbf{}& \\\hline
      1 &         GALE &    93   &  46 & \quart{41}{25}{50}{0} \\
      1 &        SWAY 4 &    102  &  15 & \quart{50}{8}{55}{0} \\
    \hline  2 &        SWAY 2 &    126  &  39 & \quart{53}{21}{68}{0} \\
      2 &       NSGA-II &    135  &  16  & \quart{65}{9}{72}{0} \\
      2 &        SPEA 2 &    143   &  10  & \quart{74}{5}{77}{0} \\
    \hline 
    \rowcolor{lightgray}\arrayrulecolor{lightgray}
    \textbf{Pom3c}  & \textbf{} & \textbf{} & \textbf{}& \\\hline
      1 &        SWAY 4 &    75   &  7 & \quart{44}{4}{45}{0} \\
      1 &        SWAY 2 &    84   &  25  & \quart{40}{16}{50}{0} \\
    \hline  2 &         GALE &    99   &  10  & \quart{57}{7}{59}{0} \\
    \hline  3 &       NSGA-II &    112  &  24 & \quart{60}{15}{66}{0} \\
      3 &        SPEA 2 &    128   &  20 & \quart{67}{12}{77}{0} \\
    
      \end{tabular}} 
    \begin{center}a. Spread  ({\em less} is {\em better})\end{center}
    \end{minipage}~~~~~~~~~~~~
    \begin{minipage}{0.5\paperwidth}
    {\small   
    \begin{tabular}{{l@{~~~~}l@{~~~~}r@{~~~~}r@{~~}c@{}r}}
    
     \multicolumn{1}{l}{Rank}& \textbf{using} & \textbf{med.} & \textbf{IQR} & \\ 
    
    \rowcolor{lightgray}\arrayrulecolor{lightgray}
    \textbf{Pom3a} & \textbf{} & \textbf{} & \textbf{} & \\\hline
      1 &        SPEA 2 &    106   &  0  & \quartr{79}{0}{79}{0} \\
      1 &       NSGA-II &    106   &  0   & \quartr{79}{0}{79}{0} \\
    \hline  2 &        SWAY 4 &    104   &  0   & \quartr{78}{0}{78}{0} \\
    \hline  3 &        SWAY 2 &    102  &  3  & \quartr{75}{3}{76}{0} \\
    \hline  4 &         GALE &    100  &  2 & \quartr{74}{1}{75}{0} \\
    \hline  
    \rowcolor{lightgray}\arrayrulecolor{lightgray}
    \textbf{Pom3b} & \textbf{} & \textbf{} & \textbf{}& \\\hline
    
      1 &       NSGA-II &    202  &  27  & \quart{69}{10}{77}{0} \\
      1 &        SPEA 2 &    184   &  8 & \quart{69}{3}{70}{0} \\
    \hline  2 &        SWAY 4 &    137   &  10 & \quart{50}{4}{52}{0} \\
    \hline 
     3 &         GALE &    95  &  23  & \quart{32}{9}{36}{0} \\
     3 &        SWAY 2 &    91    &  12  & \quart{31}{4}{35}{0} \\
     
    \hline  
    \rowcolor{lightgray}\arrayrulecolor{lightgray}
    \textbf{Pom3c} & \textbf{} & \textbf{} & \textbf{}& \\\hline
      1 &       NSGA-II &    105  &  0 & \quartr{79}{0}{79}{0} \\
      1 &        SPEA 2 &    105  &  0  & \quartr{79}{0}{79}{0} \\
    \hline  2 &        SWAY 4 &    103  &  1 & \quartr{77}{1}{78}{0} \\
    \hline 
    3 &        SWAY 2 &    101  &  1 & \quartr{76}{1}{76}{0} \\
    3 &         GALE &    100  &  1 & \quartr{75}{1}{76}{0} \\

      \end{tabular}} 
          \begin{center}b. Hypervolume  ({\em more} is {\em better})\end{center}

      \end{minipage}

    \begin{minipage}{0.5\paperwidth}

      \pgfplotstableread{
0  4000 4000 100  13  70
1 4000 4000 100  13  70   
2 4000 4000 100  13  70
        }\dataset
        \begin{tikzpicture}
        \begin{semilogyaxis}[ybar,
                width=12cm,
                height=8cm,
                xmin=-0.5,
                xmax=2.5,
                ymin=10,
                ymax=50000,        
                ylabel={Model Evaluation \#},
                xtick=data,
                xticklabels = {
                    \strut POM3A,
                    \strut POM3B,
                    \strut POM3C,
                },
                major x tick style = {opacity=0},
                minor x tick num = 1,
                minor tick length=0.5ex,
                legend entries={NSGA-II, SPEA2, GALE, SWAY2, SWAY4},
                legend columns=5,
                legend style={draw=none,nodes={inner sep=3pt}},
                ]
        \addplot[draw=black,fill=black] table[x index=0,y index=1] \dataset; 
        \addplot[draw=black,fill=black!70] table[x index=0,y index=2] \dataset; 
        \addplot[draw=black,fill=black!50] table[x index=0,y index=3] \dataset; 
        \addplot[draw=black,fill=black!30] table[x index=0,y index=4] \dataset; 
        \addplot[draw=black,fill=black!10] table[x index=0,y index=5] \dataset; 

\end{semilogyaxis}
\end{tikzpicture}
\begin{center}c. Median evaluations\end{center}
\end{minipage}

}
 ~\\
 ~\\
\hrule
 ~\\
 ~\\
  \resizebox{0.975\textwidth}{!}{
\centering
    \begin{minipage}{0.5\paperwidth}
    
    {\small   \begin{tabular}{l@{~~~~}l@{~~~~}r@{~~~~}r@{~~}c@{}r}
    
      \multicolumn{1}{l}{\textbf{Rank}}& \textbf{using} & \textbf{med.} & \textbf{IQR} & \\ 
          \rowcolor{lightgray}\arrayrulecolor{lightgray}
    \textbf{Flight}  & \textbf{} & \textbf{} & \textbf{}& \\\hline
      1 &         GALE &    100   &  15 & \quart{48}{8}{52}{0} \\
    \hline  2 &        SWAY 2 &    130   &  31 & \quart{56}{16}{67}{0} \\
      2 &        SWAY 4 &    131  &  24  & \quart{67}{12}{68}{0} \\
      2 &       NSGA-II &    143   &  12  & \quart{71}{7}{74}{0} \\
      2 &        SPEA 2 &    144   &  12 & \quart{73}{6}{74}{0} \\
    \hline  
    \rowcolor{lightgray}\arrayrulecolor{lightgray}
    \textbf{Ground}  & \textbf{} & \textbf{} & \textbf{}& \\\hline
      1 &         GALE &    100   &  48  & \quart{26}{20}{42}{0} \\
      1 &        SWAY 2 &    126   &  37  & \quart{39}{16}{53}{0} \\
    \hline  2 &        SWAY 4 &    169   &  15 & \quart{67}{6}{71}{0} \\
      2 &       NSGA-II &    169   &  23  & \quart{68}{9}{71}{0} \\
      2 &        SPEA 2 &    180   &  20 & \quart{71}{8}{76}{0} \\
    \hline  
    \rowcolor{lightgray}\arrayrulecolor{lightgray}
    \textbf{OSP}  & \textbf{} & \textbf{} & \textbf{}& \\\hline
      1 &        SWAY 2 &    88  &  20  & \quart{38}{11}{44}{0} \\
      1 &         GALE &    100  &  31  & \quart{44}{16}{51}{0} \\
    \hline  2 &        SWAY 4 &    152  &  17  & \quart{70}{9}{77}{0} \\
      2 &        SPEA 2 &    152   &  9 & \quart{75}{4}{77}{0} \\
      2 &       NSGA-II &    155    & 9 & \quart{75}{4}{79}{0} \\
    \hline  
    \rowcolor{lightgray}\arrayrulecolor{lightgray}
    \textbf{OSP2}  & \textbf{} & \textbf{} & \textbf{}& \\\hline
      1 &         GALE &    100   &  85  & \quart{12}{19}{21}{0} \\
    \hline  2 &        SWAY 2 &    210   &  115 & \quart{38}{25}{45}{0} \\
      2 &        SWAY 4 &    281   &  17 & \quart{58}{4}{60}{0} \\
    \hline  3 &        SPEA 2 &    311  &  34 & \quart{63}{7}{66}{0} \\
    \hline  4 &       NSGA-II &    355  &  38 & \quart{71}{8}{76}{0} \\
    
      \end{tabular}} 
    \begin{center}a. Spread  ({\em less} is {\em better})\end{center}
    \end{minipage}~~~~~~~~~~~~
    \begin{minipage}{0.5\paperwidth}
    {\small   
    \begin{tabular}{{l@{~~~~}l@{~~~~}r@{~~~~}r@{~~}c@{}r}}
    
     \multicolumn{1}{l}{Rank}& \textbf{using} & \textbf{med.} & \textbf{IQR} & \\ 
    \rowcolor{lightgray}\arrayrulecolor{lightgray}
    \textbf{Flight}& \textbf{} & \textbf{} & \textbf{}& \\\hline
      1 &       NSGA-II &    147   &  1 & \quartr{79}{0}{79}{0} \\
      1 &        SPEA 2 &    147   &  1 & \quartr{79}{0}{79}{0} \\
    \hline  2 &        SWAY 4 &    140  &  5  & \quartr{73}{3}{75}{0} \\
    \hline  3 &         GALE &    100  &  3 & \quart{54}{1}{54}{0} \\
      3 &        SWAY 2 &    100  &  9  & \quart{54}{5}{54}{0} \\
    \hline  
    \rowcolor{lightgray}\arrayrulecolor{lightgray}
    \textbf{Ground} & \textbf{} & \textbf{} & \textbf{}& \\\hline
      1 &       NSGA-II &    205  &  1 & \quartr{79}{0}{79}{0} \\
      1 &        SPEA 2 &    204   &  1 & \quartr{79}{0}{79}{0} \\
    \hline  2 &        SWAY 4 &    196  &  11 & \quartr{74}{4}{76}{0} \\
    \hline 
    
      3 &        SWAY 2 &    157  &  52 & \quart{47}{20}{61}{0} \\
      3 &         GALE &    100   &  17  & \quart{32}{7}{38}{0} \\
    \hline  
    \rowcolor{lightgray}\arrayrulecolor{lightgray}
    \textbf{OSP} & \textbf{} & \textbf{} & \textbf{}& \\\hline
      1 &       NSGA-II &    261  &  1  & \quartr{79}{0}{79}{0} \\
      1 &        SPEA 2 &    261  &  1 & \quartr{79}{0}{79}{0} \\
    \hline  2 &        SWAY 4 &    245   &  8  & \quartr{72}{3}{75}{0} \\
    \hline 
      3 &        SWAY 2 &    148   &  70  & \quart{30}{22}{45}{0} \\
      3 &         GALE &    100   &  0  & \quart{30}{0}{30}{0} \\
    \hline  
    \rowcolor{lightgray}\arrayrulecolor{lightgray}
    \textbf{OSP2}& \textbf{} & \textbf{} & \textbf{}& \\\hline
      1 &       NSGA-II &    171  &  1 & \quartr{79}{0}{79}{0} \\
      1 &        SPEA 2 &    171   &  0  & \quartr{79}{0}{79}{0} \\
    \hline  2 &        SWAY 4 &    163  &  10  & \quartr{72}{5}{76}{0} \\
    \hline 
    3 &        SWAY 2 &    101  &  29  & \quart{45}{14}{47}{0} \\
    3 &         GALE &    100   &  46  & \quart{46}{22}{46}{0} \\
      \end{tabular}} 
          \begin{center}b. Hypervolume  ({\em more} is {\em better})\end{center}

      \end{minipage}

    \begin{minipage}{0.5\paperwidth}

      \pgfplotstableread{
0  4000 4000 100  11  70
1 4000 4000 100  13  73   
2 4000 4000 100  12  60
3 4000 4000 100  15  62

        }\dataset
        \begin{tikzpicture}
        \begin{semilogyaxis}[ybar,
                width=12cm,
                height=10cm,
                xmin=-0.5,
                xmax=3.5,
                ymin=10,
                ymax=50000,        
                ylabel={Model Evaluation \#},
                xtick=data,
                xticklabels = {
                    \strut XOMO-Flight,
                    \strut XOMO-Ground,
                    \strut XOMO-OSP,
                    \strut XOMO-OPS2
                },
                major x tick style = {opacity=0},
                minor x tick num = 1,
                minor tick length=0.5ex,
                legend entries={NSGA-II, SPEA2, NSGA-IISC, SPEA2SC, SWAY4},
                legend columns=5,
                legend style={draw=none,nodes={inner sep=3pt}},
                ]
        \addplot[draw=black,fill=black] table[x index=0,y index=1] \dataset; 
        \addplot[draw=black,fill=black!70] table[x index=0,y index=2] \dataset; 
        \addplot[draw=black,fill=black!50] table[x index=0,y index=3] \dataset; 
        \addplot[draw=black,fill=black!30] table[x index=0,y index=4] \dataset; 
        \addplot[draw=black,fill=black!10] table[x index=0,y index=5] \dataset; 
\end{semilogyaxis}
\end{tikzpicture}
\begin{center}c. Median evaluations\end{center}
\end{minipage}}
  ~\\
 ~\\

 \caption{
 Spread and hypervolumes seen in 20 repeats.
 {\bf Med} is the 50th percentile and {\bf IQR} is the {\em inter-quartile range}; i.e. 75th-25th percentile. 
    Lines with a dot in the middle (e.g.\protect\quartex{3}{13}{13}{0})
   show the median as a  round dot within the IQR (and if the IQR is vanishingly small, only a   round dot will be visible). All results sorted by the median value:
   spread results are sorted ascending (since {\em less} spread is {\em better}) while
   hypervolume results are sorted descending (since {\em more} hypervolume is {\em better}).
   The left-hand side columns {\bf Rank} the optimizers. The {\em Smaller} the {\bf Rank}, the {\em better} the optimizer;
   e.g.  top-left, SWAY2, GALE, SWAY4 are top ranked for spread within Pom3a  with a {\bf Rank} of ``1''.
   One row has a larger ``{\bf Rank}'' than the next if (a)~its
   median values are worse and (b)~a statistical hypothesis test concurs that the distributions in the two rows are different
   (with a non-small effect
   size).
    {\bf Rank} is computed using Scott-Knott, bootstrap 95\% confidence,
   and the A12 test (see text for details). Red dots ``\textcolor{red}{\huge$\bullet$}''
   denote median results that are ``reasonable close'' to the top-ranked result (see text for details).
 }
   
\label{fig:unc_jur}
\end{figure*}

The above considerations lead to three research questions:

\textit{RQ1: Can \sway{} perform ``as good as'' traditional MOEA in unconstrained models?}

SE has seen numerous case studies where EAs like SPEA2, NSGA-II etc. have been successfully used to solve SE problems.  The unconstrained models considered in this paper are POM and XOMO. Both the models have  continuous decision space and have no constraints. In this paper we use three measures to evaluate the effectiveness of \sway{} namely, Spread, Hypervolume and \#Evalutions.

\textit{RQ2: Can \sway{} perform ``as good as'' traditional MOEA in constrained models?}

Constrained model considered in this paper is MONRP, which unlike the unconstrained models is defined in discrete space and has higher dimensionality. Note that RQ2 is particularly essential since it can expose various sensitivities of \sway{}.

\textit{RQ3: Can \sway{} be used to boost or super-charge the performance of other MOEAs?}

Since, \sway{} uses very few evaluations, it can be potentially  used as a preprocessing step for other MOEAs for cases where function evaluation is not expensive.

\begin{figure*}

 ~\\
 ~\\
  \resizebox{0.975\textwidth}{!}{
\centering
\begin{minipage}{0.5\paperwidth}
 {\small \begin{tabular}{l@{~~~}l@{~~~}r@{~~~}r@{~~~}c}
        \arrayrulecolor{lightgray}
        \textbf{Rank} & \textbf{Treatment} & \textbf{Median} & \textbf{IQR} & \\\hline
        \rowcolor{lightgray}\arrayrulecolor{lightgray}
        \textbf{50-4-5-0-110}  & \textbf{} & \textbf{} & \textbf{}& \\\hline
        \hline  1 & SWAY4 &    105  &  14 & \quart{57}{8}{62}{0} \\
        \hline  2 &    SPEA2 &    112  &  23 & \quart{61}{14}{66}{0} \\
        \hline  3 &   NSGAII &    130  &  19 & \quart{73}{11}{76}{0} \\
        
        \hline \end{tabular}}

        {\small \begin{tabular}{l@{~~~}l@{~~~}r@{~~~}r@{~~~}c}
        \rowcolor{lightgray}\arrayrulecolor{lightgray}
        \textbf{50-4-5-0-090}  & \textbf{} & \textbf{} & \textbf{}& \\\hline
          1 & SWAY4 &    98  &  8 & \quart{38}{4}{40}{0} \\
        \hline  2 &    SPEA2 &    137  &  45 & \quart{48}{18}{56}{0} \\
        \hline  3 &   NSGAII &    167  &  34 & \quart{63}{14}{68}{0} \\
        
        \hline \end{tabular}}

        {\small \begin{tabular}{l@{~~~}l@{~~~}r@{~~~}r@{~~~}c}
        \rowcolor{lightgray}\arrayrulecolor{lightgray}
        \textbf{50-4-5-4-110}  & \textbf{} & \textbf{} & \textbf{}& \\\hline
        \hline  1 & SWAY4 &    97  &  15 & \quart{54}{9}{59}{0} \\
        \hline  2 &    SPEA2 &    113  &  11 & \quart{65}{7}{68}{0} \\
        \hline  3 &   NSGAII &    125  &  28 & \quart{65}{17}{75}{0} \\
        
        \hline \end{tabular}}

        {\small \begin{tabular}{l@{~~~}l@{~~~}r@{~~~}r@{~~~}c}
        \rowcolor{lightgray}\arrayrulecolor{lightgray}
        \textbf{50-4-5-4-090}  & \textbf{} & \textbf{} & \textbf{}& \\\hline
          1 & SWAY4 &    102  &  12 & \quart{56}{7}{60}{0} \\
        \hline  2 &    SPEA2 &    119  &  21 & \quart{65}{12}{70}{0} \\
          2 &   NSGAII &    133  &  21 & \quart{71}{12}{78}{0} \\
        
        \hline \end{tabular}}

    \begin{center}a. Spread  ({\em less} is {\em better})\end{center}
    \end{minipage}~~~~~~~~~~~~
    \begin{minipage}{0.5\paperwidth}
    {\small \begin{tabular}{l@{~~~}l@{~~~}r@{~~~}r@{~~~}c}
        \arrayrulecolor{lightgray}
        \textbf{Rank} & \textbf{Treatment} & \textbf{Median} & \textbf{IQR} & \\\hline
        \rowcolor{lightgray}\arrayrulecolor{lightgray}
        \textbf{50-4-5-0-110}  & \textbf{} & \textbf{} & \textbf{}& \\\hline
          1 &    SPEA2 &    157  &  16 & \quart{72}{7}{75}{0} \\
          1 &   NSGAII &    154  &  13 & \quart{69}{6}{73}{0} \\
        \hline  2 & SWAY4 &    111  &  6 & \quart{52}{3}{53}{0} \\
        \hline \end{tabular}}
        
        {\small \begin{tabular}{l@{~~~}l@{~~~}r@{~~~}r@{~~~}c}
        \rowcolor{lightgray}\arrayrulecolor{lightgray}
        \textbf{50-4-5-0-090}  & \textbf{} & \textbf{} & \textbf{}& \\\hline
           1 & SWAY4 &    103  &  8 & \quart{68}{5}{70}{0} \\
          1 &    SPEA2 &    105  &  20 & \quart{64}{14}{71}{0} \\
          1 &   NSGAII &    104  &  19 & \quart{67}{12}{70}{0} \\ 
        \hline \end{tabular}}
        
        {\small \begin{tabular}{l@{~~~}l@{~~~}r@{~~~}r@{~~~}c}
        \rowcolor{lightgray}\arrayrulecolor{lightgray}
        \textbf{50-4-5-4-110}  & \textbf{} & \textbf{} & \textbf{}& \\\hline
          1 & SWAY4 &    145  &  5 & \quart{77}{2}{79}{0} \\ 
        \hline  2 &    SPEA2 &    68  &  3 & \quart{36}{1}{37}{0} \\
          2 &   NSGAII &    66  &  5 & \quart{34}{3}{35}{0} \\
        \hline \end{tabular}}
        
        {\small \begin{tabular}{l@{~~~}l@{~~~}r@{~~~}r@{~~~}c}
        \rowcolor{lightgray}\arrayrulecolor{lightgray}
        \textbf{50-4-5-4-090}  & \textbf{} & \textbf{} & \textbf{}& \\\hline
        \hline  1 & SWAY4 &    73  &  3 & \quart{56}{3}{58}{0} \\
        \hline  2 &    SPEA2 &    63  &  4 & \quart{47}{4}{49}{0} \\
          2 &   NSGAII &    64  &  5 & \quart{47}{4}{50}{0} \\
        \hline \end{tabular}}

    \begin{center}b. Hypervolume ({\em more} is {\em better})\end{center}
    \end{minipage}

\begin{minipage}{0.5\paperwidth}

      \pgfplotstableread{
0  2000 2000 73
1 2000 2000 73
2 2000  2000 69
3 2000  2000 76
        }\dataset
        \begin{tikzpicture}
        \begin{semilogyaxis}[ybar,
                width=12cm,
                height=8cm,
                xmin=-0.5,
                xmax=3.5,
                ymin=10,
                ymax=50000,        
                ylabel={Model Evaluation \#},
                xtick=data,
                xticklabels = {
                    \strut 50-4-5-0-110,
                    \strut 50-4-5-0-090,
                    \strut 50-4-5-4-110,
                    \strut 50-4-5-4-090
                },
                major x tick style = {opacity=0},
                minor x tick num = 1,
                minor tick length=0.5ex,
                legend entries={NSGA-II, SPEA2, SWAY4},
                legend columns=5,
                legend style={draw=none,nodes={inner sep=3pt}},
                ]
        \addplot[draw=black,fill=black] table[x index=0,y index=1] \dataset; 
        \addplot[draw=black,fill=black!70] table[x index=0,y index=2] \dataset; 
        \addplot[draw=black,fill=black!10] table[x index=0,y index=3] \dataset; 

\end{semilogyaxis}
\end{tikzpicture}
\begin{center}c. Median evaluations\end{center}
\end{minipage}

}

 \caption{
 Spread and hypervolumes seen in 20 repeats for the constrained problem (MONRP).
 Results are presented in the same format as Figure~\ref{fig:unc_jur}.
 }
   
\label{fig:c_jur}
\end{figure*}
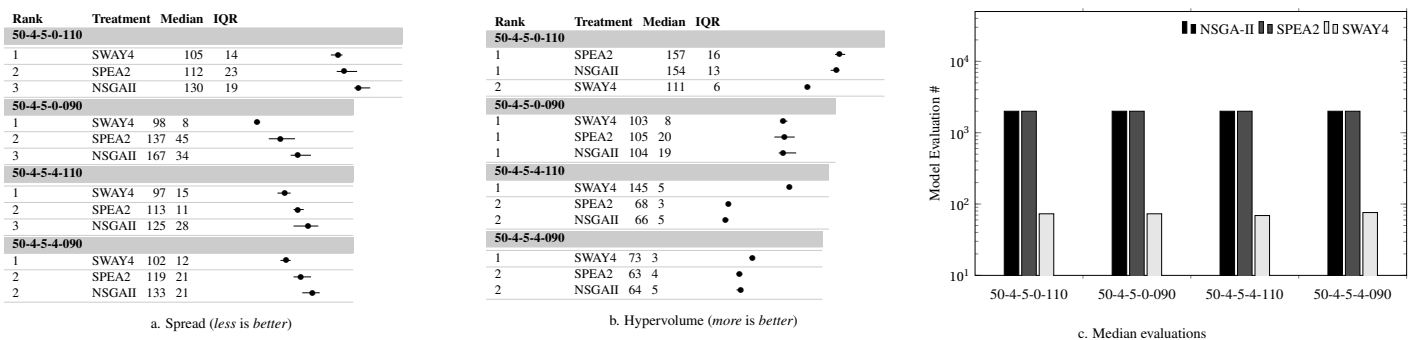


\subsection{Experimental Setup}

In the following, we compare EAs to SWAY for 20 repeats of our two unconstrained models through
our various scenarios. All the optimizers use the population size
recommended by their original authors; i.e. $n=100$.
But, to test the effects of increased sample, we run two
versions of SWAY:
\bi
\item SWAY2: builds an initial population of size $10^2=100$. 
\item SWAY4: builds an initial population of size $10^4=10,000$.
\ei
One design choice in this experiment was the evaluation budget for each optimizer:
\bi
\item
If we increase number of iterations in EA to a very large number, that would bias the comparison towards
EAs since   better optimizations might be found just by
blind luck~(albeit at infinite cost). 
\item
Conversely, if we restrict EAs to the number of evaluations made by (say) SWAY4 then
that would unfairly bias the comparison towards
SWAY since that would allow only a generation or two of EA.
\ei

\rv{
Consequently, to compare SWAY and EA, following the practice
of~\cite{galepaper,krall15:hms}, we set maximum number
of evaluations to 2000.}

\section{Results}

\subsection{RQ1 -- Can \sway{} perform ``as good as'' traditional MOEA in unconstrained models?}

Figure~\ref{fig:unc_jur} shows results obtained by all the optimizers, compared to the results obtained using SWAY techniques.
The figure shows the median ({\bf med.}) and inter quartile range ({\bf IQR} 75th-25th value) 
for all the optimizers and SWAY techniques. Horizontal quartile plots show the median as a round dot within
the inter-quartile range. 
In the figure, an optimizer's score is ranked 1~({\bf Rank}=1) 
if other optimizers have (a)~worse medians;
and (b)~the other distributions are significantly different~(computed via Scott-Knott
and bootstrapping); and (c)~differences are
not a small effect (computed via A12).

Figure~\ref{fig:unc_jur}(a) shows the spread results and
can be summarized as:
the spreads found by standard EAs (NSGA-II and SPEA2) were always  ranked last in
all scenarios.  That is, for these scenarios and models,
to achieve a good distribution of results, it is better
to sample than evolve.

Figure~\ref{fig:unc_jur}(b) shows the hypervolume results and
can be summarized as:
GALE and SWAY2 were always ranked last in all scenarios.  That is,
for these scenarios and models,
to find best optimization solutions, it is insufficient to explore
just a few evaluations of a small population (e.g. the 100 instances explored
by SWAY2 and GALE).

\fig{unc_jur}(c) shows the number of evaluations for our optimizers. Note that:
\bi
\item GALE requires more evaluations than SWAY since SWAY terminates after one generation
while GALE runs for multiple evaluations.
\item
Even though
SWAY4 explores 100 times the population of SWAY2, it only has to evaluate logarithmically
more individuals- so the total number of extra evaluations for SWAY4 may only increase 2 to 4 times
from SWAY2.
\item
The standard optimizers (NSGA-II and SPEA2) require orders of magnitude more evaluations.
This is because these optimizers evaluate all $n$ members of each population,
GALE and SWAY, on the other hand, 
only evaluate  $2\mathit{log}{n}$ members.
\ei

Figure~\ref{fig:unc_jur}(a) and Figure~\ref{fig:unc_jur}(b)  made a case against SWAY2, GALE, and our EAs.
As to SWAY4, we note that SWAY4's spread is never worse than standard EAs (and sometimes it is even best: see the Pom3s spread results).
As to the SWAY4 hypervolume results, in one case (Pom3b), SWAY4 is clearly
inferior to standard EAs (NSGA-II and SPEA2). But in all the other results,
SWAY4 is an interesting option. Often it is ranked second after EAs
but   those statistical rankings do not always
pass a ``reasonableness'' test. Consider the hypervolumes achieved 
in Pom3a: 106,106,104,102,100
where  the best hypervolume (of 106) came from SPEA2 while SWAY4  generated
very similar  hypervolumes of 104.
Our statistical tests
divide optimizers with median values of 
106,106,104,102,100  into four {\bf Ranks}: which may not be ``reasonable''.
As pragmatic engineers, we are hard-pressed to recommend   evaluating a very slow model
$2,000$ times to achieve a  hypervolume of 106 (using SPEA2) when 50 evaluations of SWAY4
would achieve a hypervolume of 104. In Figure~\ref{fig:unc_jur}, we mark all the results that we think are
 ``reasonable close'' to the top-ranked result with a red ``\textcolor{red}{\huge$\bullet$}''
dot.  SWAY4 is always marked as ``reasonable close'' to
the EAs.

We acknowledge that  the use of the  ``reasonableness'' measure in the last
paragraph is somewhat   subjective assessment. Also,
for some ultra-mission critical domains, it might indeed be required to select
optimizers that generate hypervolumes that are $\frac{106 - 104}{104} = 2\%$ better
than anything else.  However, we suspect that many engineers would
gladly use a method that is 50 times faster and delivers (very nearly) the same results.

\begin{figure*}[!th]
 \resizebox{0.975\textwidth}{!}{
\centering
    \begin{minipage}{0.5\paperwidth}
    
    {\small   \begin{tabular}{l@{~~~~}l@{~~~~}r@{~~~~}r@{~~}c@{}r}
    
      \multicolumn{1}{l}{\textbf{Rank}}& \textbf{using} & \textbf{med.} & \textbf{IQR} & \\ 
      
    \rowcolor{lightgray}\arrayrulecolor{lightgray}
    \textbf{Pom3a} & \textbf{} & \textbf{} & \textbf{} & \\\hline
  1 &   NSGAII &    151  &  22.97 & \quart{52}{22}{61}{-90} \\
  1 &    SPEA2 &    152  &  24.19 & \quart{57}{22}{62}{-90} \\
  1 & NSGAIISC &    153  &  20.33 & \quart{56}{19}{63}{-90} \\
  1 & SPEA2SC &    156  &  24.93 & \quart{48}{23}{66}{-90} \\
    \hline  
    \rowcolor{lightgray}\arrayrulecolor{lightgray}
    \textbf{Pom3b} & \textbf{} & \textbf{} & \textbf{}& \\\hline
    1 & SPEA2SC &    130  &  12.86 & \quart{49}{14}{58}{-97} \\
  1 &    SPEA2 &    131  &  20.0 & \quart{51}{21}{59}{-97} \\
  1 &   NSGAII &    135  &  17.27 & \quart{52}{18}{64}{-97} \\
\hline 
  2 & NSGAIISC &    139  &  18.17 & \quart{58}{19}{70}{-97} \\
    \hline 
    \rowcolor{lightgray}\arrayrulecolor{lightgray}
    \textbf{Pom3c}  & \textbf{} & \textbf{} & \textbf{}& \\\hline
   1 & NSGAIISC &    109  &  24.23 & \quart{34}{31}{43}{-120} \\
  1 &   NSGAII &    112  &  24.33 & \quart{31}{31}{46}{-120} \\
\hline  
  2 &    SPEA2 &    121  &  14.64 & \quart{52}{18}{59}{-120} \\
  2 & SPEA2SC &    121  &  9.58 & \quart{56}{12}{60}{-120} \\

      \end{tabular}} 
    \begin{center}a. Spread  ({\em less} is {\em better})\end{center}
    \end{minipage}~~~~~~~~~~~~
    \begin{minipage}{0.5\paperwidth}
    {\small   
    \begin{tabular}{{l@{~~~~}l@{~~~~}r@{~~~~}r@{~~}c@{}r}}
    
     \multicolumn{1}{l}{Rank}& \textbf{using} & \textbf{med.} & \textbf{IQR} & \\ 
    
    \rowcolor{lightgray}\arrayrulecolor{lightgray}
    \textbf{Pom3a} & \textbf{} & \textbf{} & \textbf{} & \\\hline
  1 &    SPEA2 &    106  &  0.43 & \quart{75}{2}{76}{-526} \\
  1 & NSGAIISC &    106  &  0.34 & \quart{75}{2}{76}{-526} \\
  1 &   NSGAII &    106  &  0.54 & \quart{75}{3}{77}{-526} \\
  1 & SPEA2SC &    106  &  0.57 & \quart{74}{4}{76}{-526} \\
    \hline  
    \rowcolor{lightgray}\arrayrulecolor{lightgray}
    \textbf{Pom3b} & \textbf{} & \textbf{} & \textbf{}& \\\hline
  1 &    SPEA2 &    202  &  0.74 & \quart{75}{3}{76}{-379} \\
  1 & NSGAIISC &    202  &  0.93 & \quart{75}{4}{76}{-379} \\
  1 &   NSGAII &    202  &  0.55 & \quart{75}{2}{76}{-379} \\
  1 & SPEA2SC &    202  &  1.19 & \quart{74}{5}{76}{-379} \\
    \hline  
    \rowcolor{lightgray}\arrayrulecolor{lightgray}
    \textbf{Pom3c} & \textbf{} & \textbf{} & \textbf{}& \\\hline
  1 &   NSGAII &    105  &  3.04 & \quart{73}{4}{75}{-114} \\
  1 & SPEA2SC &    105  &  4.39 & \quart{71}{7}{74}{-114} \\
\hline  2 &    SPEA2 &    104  &  3.75 & \quart{72}{5}{73}{-114} \\
  2 & NSGAIISC &    104  &  6.02 & \quart{70}{9}{73}{-114} \\

      \end{tabular}} 
          \begin{center}b. Hypervolume  ({\em more} is {\em better})\end{center}

      \end{minipage}

    \begin{minipage}{0.5\paperwidth}

      \pgfplotstableread{
0  4000 4000 4094  4094 
1 4000 4000 4090  4090   
2 4000 4000 4096  4096 
        }\dataset
        \begin{tikzpicture}
        \begin{semilogyaxis}[ybar,
                width=12cm,
                height=8cm,
                xmin=-0.5,
                xmax=2.5,
                ymin=10,
                ymax=50000,        
                ylabel={Model Evaluation \#},
                xtick=data,
                xticklabels = {
                    \strut POM3A,
                    \strut POM3B,
                    \strut POM3C
                },
                major x tick style = {opacity=0},
                minor x tick num = 1,
                minor tick length=0.5ex,
                legend entries={NSGA-II, SPEA2, NSGA-IISC, SPEA2SC},
                legend columns=4,
                legend style={draw=none,nodes={inner sep=3pt}},
                ]
        \addplot[draw=black,fill=black] table[x index=0,y index=1] \dataset; 
        \addplot[draw=black,fill=black!70] table[x index=0,y index=2] \dataset; 
        \addplot[draw=black,fill=black!50] table[x index=0,y index=3] \dataset; 
        \addplot[draw=black,fill=black!30] table[x index=0,y index=4] \dataset;

\end{semilogyaxis}
\end{tikzpicture}
\begin{center}c. Median evaluations\end{center}
\end{minipage}

}
 ~\\
 ~\\
\hrule
 ~\\
 ~\\
  \resizebox{0.975\textwidth}{!}{
\centering
    \begin{minipage}{0.5\paperwidth}
    
    {\small   \begin{tabular}{l@{~~~~}l@{~~~~}r@{~~~~}r@{~~}c@{}r}
    
      \multicolumn{1}{l}{\textbf{Rank}}& \textbf{using} & \textbf{med.} & \textbf{IQR} & \\ 
          \rowcolor{lightgray}\arrayrulecolor{lightgray}
    \textbf{Flight}  & \textbf{} & \textbf{} & \textbf{}& \\\hline
   1 &    SPEA2 &    129  &  7.77 & \quart{53}{8}{55}{-99} \\
  1 & SPEA2SC&    134  &  10.02 & \quart{56}{10}{60}{-99} \\
\hline  3 & NSGAIISC &    137  &  20.89 & \quart{54}{21}{64}{-99} \\
  2 &   NSGAII &    143  &  22.42 & \quart{56}{23}{70}{-99} \\
    \hline  
    \rowcolor{lightgray}\arrayrulecolor{lightgray}
    \textbf{Ground}  & \textbf{} & \textbf{} & \textbf{}& \\\hline
  1 &   NSGAII &    169  &  13.32 & \quart{52}{13}{60}{-89} \\
  2 &    SPEA2 &    170  &  14.51 & \quart{52}{15}{61}{-89} \\
  2 & SPEA2SC&    170  &  6.28 & \quart{58}{6}{61}{-89} \\
\hline  3 & NSGAIISC &    178  &  15.12 & \quart{58}{15}{68}{-89} \\
    \hline  
    \rowcolor{lightgray}\arrayrulecolor{lightgray}
    \textbf{OSP}  & \textbf{} & \textbf{} & \textbf{}& \\\hline
  1 & SPEA2SC&    148  &  10.57 & \quart{57}{11}{61}{-92} \\
\hline  2 & NSGAIISC &    153  &  13.05 & \quart{61}{13}{65}{-92} \\
  2 &    SPEA2 &    155  &  16.13 & \quart{58}{17}{67}{-92} \\
  2 &   NSGAII &    155  &  16.62 & \quart{56}{17}{68}{-92} \\
    \hline  
    \rowcolor{lightgray}\arrayrulecolor{lightgray}
    \textbf{OSP2}  & \textbf{} & \textbf{} & \textbf{}& \\
\hline  1 &    SPEA2 &    344  &  12.16 & \quart{56}{13}{62}{-99} \\
  1 & SPEA2SC&    344  &  9.62 & \quart{60}{10}{63}{-99} \\
\hline  2 &   NSGAII &    355  &  19.43 & \quart{58}{21}{68}{-99} \\
  2 & NSGAIISC&    362  &  14.97 & \quart{63}{16}{71}{-99} \\
    
      \end{tabular}} 
    \begin{center}d. Spread  ({\em less} is {\em better})\end{center}
    \end{minipage}~~~~~~~~~~~~
    \begin{minipage}{0.5\paperwidth}
    {\small   
    \begin{tabular}{{l@{~~~~}l@{~~~~}r@{~~~~}r@{~~}c@{}r}}
    
     \multicolumn{1}{l}{Rank}& \textbf{using} & \textbf{med.} & \textbf{IQR} & \\ 
    \rowcolor{lightgray}\arrayrulecolor{lightgray}
    \textbf{Flight}& \textbf{} & \textbf{} & \textbf{}& \\\hline
  1 &    SPEA2 &    146  &  0.97 & \quart{73}{5}{75}{-479} \\
  1 & NSGAIISC &    147  &  0.74 & \quart{73}{4}{75}{-479} \\
  1 &   NSGAII &    147  &  1.31 & \quart{73}{6}{75}{-479} \\
  1 & SPEA2SC &    147  &  1.25 & \quart{73}{6}{77}{-479} \\
    \hline  
    \rowcolor{lightgray}\arrayrulecolor{lightgray}
    \textbf{Ground} & \textbf{} & \textbf{} & \textbf{}& \\\hline
  1 &    SPEA2 &    205  &  1.19 & \quart{74}{5}{76}{-394} \\
  1 &   NSGAII &    205  &  0.94 & \quart{74}{4}{76}{-394} \\
\hline  
  2 & NSGAIISC &    203  &  0.85 & \quart{72}{4}{74}{-394} \\
  2 & SPEA2SC &    203  &  0.89 & \quart{73}{4}{74}{-394} \\ \\
    \hline  
    \rowcolor{lightgray}\arrayrulecolor{lightgray}
    \textbf{OSP} & \textbf{} & \textbf{} & \textbf{}& \\\hline
  1 &    SPEA2 &    261  &  1.22 & \quart{71}{6}{73}{-440} \\
  1 & NSGAIISC &    261  &  1.0 & \quart{72}{5}{73}{-440} \\
  1 &   NSGAII &    261  &  1.16 & \quart{71}{6}{74}{-440} \\
  1 & SPEA2SC &    261  &  1.51 & \quart{71}{8}{75}{-440} \\
    \hline  
    \rowcolor{lightgray}\arrayrulecolor{lightgray}
    \textbf{OSP2}& \textbf{} & \textbf{} & \textbf{}& \\\hline
  1 &    SPEA2 &    171  &  1.37 & \quart{72}{7}{75}{-490} \\
  1 &   NSGAII &    171  &  1.03 & \quart{71}{6}{74}{-490} \\
  1 & SPEA2SC &    171  &  0.95 & \quart{73}{5}{75}{-490} \\
  1 & NSGAIISC &    171  &  0.91 & \quart{73}{5}{75}{-490} \\
      \end{tabular}} 
          \begin{center}e. Hypervolume  ({\em more} is {\em better})\end{center}

      \end{minipage}

    \begin{minipage}{0.5\paperwidth}

      \pgfplotstableread{
0  4000 4000 4094  4094 
1 4000 4000 4090  4090     
2 4000 4000 4096  4096  
3 4000 4000 4103 4103 

        }\dataset
        \begin{tikzpicture}
        \begin{semilogyaxis}[ybar,
                width=12cm,
                height=10cm,
                xmin=-0.5,
                xmax=3.5,
                ymin=10,
                ymax=50000,        
                ylabel={Model Evaluation \#},
                xtick=data,
                xticklabels = {
                    \strut XOMO-Flight,
                    \strut XOMO-Ground,
                    \strut XOMO-OSP,
                    \strut XOMO-OPS2
                },
                major x tick style = {opacity=0},
                minor x tick num = 1,
                minor tick length=0.5ex,
                legend entries={NSGA-II, SPEA2, NSGA-IISC, SPEA2SC},
                legend columns=4,
                legend style={draw=none,nodes={inner sep=3pt}},
                ]
        \addplot[draw=black,fill=black] table[x index=0,y index=1] \dataset; 
        \addplot[draw=black,fill=black!70] table[x index=0,y index=2] \dataset; 
        \addplot[draw=black,fill=black!50] table[x index=0,y index=3] \dataset; 
        \addplot[draw=black,fill=black!30] table[x index=0,y index=4] \dataset; 
\end{semilogyaxis}
\end{tikzpicture}
\begin{center}f. Median evaluations\end{center}
\end{minipage}}
  ~\\
 ~\\
\hrule
 ~\\
 ~\\
  \resizebox{0.975\textwidth}{!}{
\centering
\begin{minipage}{0.5\paperwidth}
 {\small \begin{tabular}{l@{~~~}l@{~~~}r@{~~~}r@{~~~}c}
        \arrayrulecolor{lightgray}
        \textbf{Rank} & \textbf{Treatment} & \textbf{Median} & \textbf{IQR} & \\\hline
        \rowcolor{lightgray}\arrayrulecolor{lightgray}
        \textbf{50-4-5-0-110}  & \textbf{} & \textbf{} & \textbf{}& \\\hline
        \hline  1 &    SPEA2 &    127  &  0.0 & \quart{68}{0}{68}{0} \\
        \hline  2 &   NSGAII &    130  &  0.0 & \quart{70}{0}{70}{0} \\
        \hline  3 &  SPEA2SC &    137  &  0.0 & \quart{74}{0}{74}{0} \\
        \hline  4 & NSGAIISC &    148  &  0.0 & \quart{79}{0}{79}{0} \\
        
        \hline \end{tabular}}

        {\small \begin{tabular}{l@{~~~}l@{~~~}r@{~~~}r@{~~~}c}
        \rowcolor{lightgray}\arrayrulecolor{lightgray}
        \textbf{50-4-5-0-090}  & \textbf{} & \textbf{} & \textbf{}&
\\\hline  1 &        SPEA2 &    140  &  23.18 & \quart{48}{11}{54}{0} \\
\hline  2 &      SPEA2SC &    158  &  30.77 & \quart{52}{14}{61}{0} \\
\hline  3 &       NSGAII &    167  &  30.74 & \quart{61}{14}{65}{0} \\
  3 &     NSGAIISC &    183 &  37.95 & \quart{62}{17}{71}{0} \\
        
        \hline \end{tabular}}

        {\small \begin{tabular}{l@{~~~}l@{~~~}r@{~~~}r@{~~~}c}
        \rowcolor{lightgray}\arrayrulecolor{lightgray}
        \textbf{50-4-5-4-110}  & \textbf{} & \textbf{} & \textbf{}& 
\\\hline  1 &        SPEA2 &    102  &  11.8 & \quart{45}{5}{48}{0} \\
\hline  2 &      SPEA2SC &    114  &  22.04 & \quart{49}{10}{53}{0} \\
\hline  3 &       NSGAII &    125  &  24.87 & \quart{54}{11}{58}{0} \\
  3 &     NSGAIISC &    145  &  45.14 & \quart{59}{20}{68}{0} \\
        
        \hline \end{tabular}}

        {\small \begin{tabular}{l@{~~~}l@{~~~}r@{~~~}r@{~~~}c}
        \rowcolor{lightgray}\arrayrulecolor{lightgray}
        \textbf{50-4-5-4-090}  & \textbf{} & \textbf{} & \textbf{}& \\\hline
    1 &        SPEA2 &    117  &  29.07 & \quart{48}{13}{56}{0} \\
  1 &      SPEA2SC &    122  &  20.85 & \quart{53}{9}{58}{0} \\
\hline  2 &       NSGAII &    133  &  32.41 & \quart{59}{14}{63}{0} \\
  2 &     NSGAIISC &    149  &  39.47 & \quart{62}{17}{71}{0} \\
        
        \hline \end{tabular}}

    \begin{center}g. Spread  ({\em less} is {\em better})\end{center}
    \end{minipage}~~~~~~~~~~~~
    \begin{minipage}{0.5\paperwidth}
    {\small \begin{tabular}{l@{~~~}l@{~~~}r@{~~~}r@{~~~}c}
        \arrayrulecolor{lightgray}
        \textbf{Rank} & \textbf{Treatment} & \textbf{Median} & \textbf{IQR} & \\\hline
        \rowcolor{lightgray}\arrayrulecolor{lightgray}
        \textbf{50-4-5-0-110}  & \textbf{} & \textbf{} & \textbf{}& \\\hline
  1 &       NSGAII &    154  &  17.94 & \quart{70}{8}{73}{0} \\
  1 &        SPEA2 &    162 &  17.18 & \quart{71}{8}{77}{0} \\
\hline  2 &      SPEA2SC &    141  &  19.17 & \quart{63}{9}{67}{0} \\
  2 &     NSGAIISC &    139  &  12.51 & \quart{63}{6}{66}{0} \\
        \hline \end{tabular}}
        
        {\small \begin{tabular}{l@{~~~}l@{~~~}r@{~~~}r@{~~~}c}
        \rowcolor{lightgray}\arrayrulecolor{lightgray}
        \textbf{50-4-5-0-090}  & \textbf{} & \textbf{} & \textbf{}& \\\hline
  1 &        SPEA2 &    103  &  22.4 & \quart{73}{6}{76}{0} \\
  1 &       NSGAII &    104  &  15.67 & \quart{68}{4}{69}{0} \\
  1 &      SPEA2SC &    102  &  15.23 & \quart{55}{13}{59}{0} \\
  1 &     NSGAIISC &    101  &  21.28 & \quart{52}{12}{59}{0} \\
        \hline \end{tabular}}
        
        {\small \begin{tabular}{l@{~~~}l@{~~~}r@{~~~}r@{~~~}c}
        \rowcolor{lightgray}\arrayrulecolor{lightgray}
        \textbf{50-4-5-4-110}  & \textbf{} & \textbf{} & \textbf{}& \\\hline
        1 &        SPEA2 &    71  &  6.28 & \quart{67}{5}{69}{0} \\
\hline  2 &       NSGAII &    66  &  10.12 & \quart{59}{8}{64}{0} \\
  2 &      SPEA2SC &    64  &  5.9 & \quart{59}{4}{62}{0} \\
\hline  3 &     NSGAIISC &    56  &  15.55 & \quart{50}{12}{53}{0} \\
        \hline \end{tabular}}
        
        {\small \begin{tabular}{l@{~~~}l@{~~~}r@{~~~}r@{~~~}c}
        \rowcolor{lightgray}\arrayrulecolor{lightgray}
        \textbf{50-4-5-4-090}  & \textbf{} & \textbf{} & \textbf{}& \\\hline
    1 &       NSGAII &    64  &  14.78 & \quart{40}{12}{48}{0} \\
  1 &        SPEA2 &    66  &  12.72 & \quart{43}{10}{49}{0} \\
  1 &      SPEA2SC &    70  &  9.87 & \quart{47}{8}{52}{0} \\
  1 &     NSGAIISC &    64  &  19.3 & \quart{35}{15}{47}{0} \\
        \hline \end{tabular}}

    \begin{center}h. Hypervolume ({\em more} is {\em better})\end{center}
    \end{minipage}

\begin{minipage}{0.5\paperwidth}

      \pgfplotstableread{
0  4000 4000 4094 4094  
1 4000 4000 4090  4090     
2 4000 4000 4096  4096  
3 4000 4000 4103 4103 
}\dataset
        \begin{tikzpicture}
        \begin{semilogyaxis}[ybar,
                width=12cm,
                height=10cm,
                xmin=-0.5,
                xmax=3.5,
                ymin=10,
                ymax=50000,        
                ylabel={Model Evaluation \#},
                xtick=data,
                xticklabels = {
                    \strut 50-4-5-0-110,
                    \strut 50-4-5-0-090,
                    \strut 50-4-5-4-110,
                    \strut 50-4-5-4-090
                },
                major x tick style = {opacity=0},
                minor x tick num = 1,
                minor tick length=0.5ex,
                legend entries={NSGA-II, SPEA2, NSGA-IISC, SPEA2SC},
                legend columns=4,
                legend style={draw=none,nodes={inner sep=3pt}},
                ]
        \addplot[draw=black,fill=black] table[x index=0,y index=1] \dataset; 
        \addplot[draw=black,fill=black!70] table[x index=0,y index=2] \dataset; 
        \addplot[draw=black,fill=black!50] table[x index=0,y index=3] \dataset; 
        \addplot[draw=black,fill=black!30] table[x index=0,y index=4] \dataset; 

\end{semilogyaxis}
\end{tikzpicture}
\begin{center}i. Median evaluations\end{center}
\end{minipage}

}

 \caption{
 \sway{} results were used to seed  the initial population of the traditional MOEA. Spread, hypervolumes and evaluation seen in 20 repeats. This figure is of the same format as Figure~\ref{fig:unc_jur}
 }
   
\label{fig:sc_fig}
\end{figure*}

\subsection{RQ2 -- Can \sway{} perform ``as good as'' traditional MOEA in constrained models?}

The results of RQ1 recommends use {\bf SWAY 4}. We also observe that:
\begin{itemize}
    \item {\bf SWAY 4} is better than {\bf SWAY 2}
    \item {\bf SWAY 4} is better than GALE
\end{itemize}
Accordingly, for the constrained models we do not consider GALE and SWAY 2. 
Further, to
compare the algorithms based on an extra parameter- ``constraint level - how constrained the problem is'' along with the parameters mentioned in \S \ref{performance_m}. 
As seen in Figure~\ref{fig:c_jur}, SPEA2 performs worse that SWAY4 in term of spread. When comparing results from Hypervolume, SPEA2  performs better than SWAY4 in only one case (MONRP-50-4-5-4-100). Please note that the MONRP-50-4-5-4-110 is the least constrained problem (among the constrained problems) considered in this paper. Hence, we cannot recommend this SPEA2 as our preferred method, for the MONRP family of problems. We observe in Figure~\ref{fig:c_jur} that:
\bi
\item
SWAY4 has the lowest spreads (lower the better) and highest hypervolume (higher the better) in most  cases. 
\item
The exception being MONRP-50-4-5-4-100, which is the least constrained along the four variants. That said we see that \sway{} is the top ranked optimizer in $\frac{4}{4}$ cases and $\frac{3}{4}$ with respect to Spread and Hypervolume respectively.
\item
Section~(c) of Figure \ref{fig:c_jur} compares the number of evaluations required by \sway{} and other optimizers. Note that \sway{} requires 20 times less evaluations compared to the standard optimizers, while performing reasonably well in other quality metrics. 
\ei
In summary, the MONRP results nearly always endorse the use of SWAY4.

\subsection{RQ3 -- Can \sway{} be used to boost or super-charge the performance of other MOEAs?}\label{rq3x}

We compare EAs which were super-charged with the results from \sway{}. Super-charging a EA means seeding the initial population with the results obtained from \sway{}. The supercharged NSGA-II and SPEA2 is called NSGA-IISC and SPEA2SC respectively.
\rv{NSGA-IISC and SPEA2SC shared exactly the same configurations as in RQ1 and RQ2 except for the initial seeds.}

The results from super-charging are shown in Figure~\ref{fig:sc_fig}. Results
that endorse super-charging would have the following form:

\bi
\item The super-charged version of the algorithm would have lower spreads or higher hypervolumes that otherwise;
\item The results from the super-charged version have a different statistical rank
(as listed in the first column of these results).
\ei
From Figure~\ref{fig:sc_fig},  there are 22 cases (each) where results
might appear that endorse super-charging for NSGA-II and SPEA2. 
\bi
\item For NSGA-II, results endorsing super-charging appear zero times;
\item For SPEA2, results endorsing super-charging appear once for spread for OSP
and once  for hypervolumes for Pom3c.
\ei

\rv{
Overall, the results endorsing super-charging occur so rarely that we cannot
recommend \sway{} as a pre-processor to other optimizers.
Even though this experiment showed that \sway{} is not a good idea for super-charging other MOEAs, 
it does suggest another, potentially promising, avenue for future SBSE
research.
A common view for evolutionary algorithms is that they find out solutions through evolution, or improvements between iterations.
Consequently, many researchers are focusing on how to customize operators between iterations so that EA's performance  can be enhanced.
However, this paper, we show that 1) \sway{}, a sampling technique, can get similar results as standard EA;
2) standard EA cannot improve significantly from \sway{}'s output.
These means  evolution is not the only method that might
be able to solve complex non-linear problems.
Future researchers might also choose to focus on
sampling methods.
}

\section{Why does this work?}\label{whywork}

In this section, we present an analysis to understand why  \sway{} achieves such low spread scores and high hypervolume scores similar to more complicated algorithms. We hypothesize that the decision space of the problem space lie on a low dimensional manifold.

\subsection{History}
\rv{The results of this paper, that \sway{},  a very simple  sampling method
works surprisingly well, is consistent with numerous other results by these authors, dating back to 2003~\cite{menzies2003data}.  Our explanation 
for this effect is that the underlying search space for SE models is 
inherently very simple due to certain emergent properties of state space
(for more details, see~\cite{menzies2007strangest}).

Recently, we have become aware of  much synergy between our prior work~\cite{menzies2003data,menzies2007strangest}
and the work of others~\cite{bettenburg2012think, deiters2013using, bettenburg2015towards, zhang2016cross}. These other results
all propose methods to learn an simpler, smaller set of  underlying dimensions
using methods like FASTMAP~\cite{faloutsos1995fastmap} or   Principal Component Analysis (PCA)~\cite{jolliffe2002principal}~\footnote{WHERE is an approximation of the first principal component}, Spectral Learning~\cite{shi2000normalized} and Random Projection~\cite{bingham2001random}.  These algorithms  use  different techniques to identify the underlying, independent/orthogonal dimensions to cluster the data points and differ with respect to the computational complexity and accuracy. We use WHERE since it computationally efficient~$O(2N)$, while still being accurate.
}

\begin{figure}
\includegraphics[width=\linewidth]{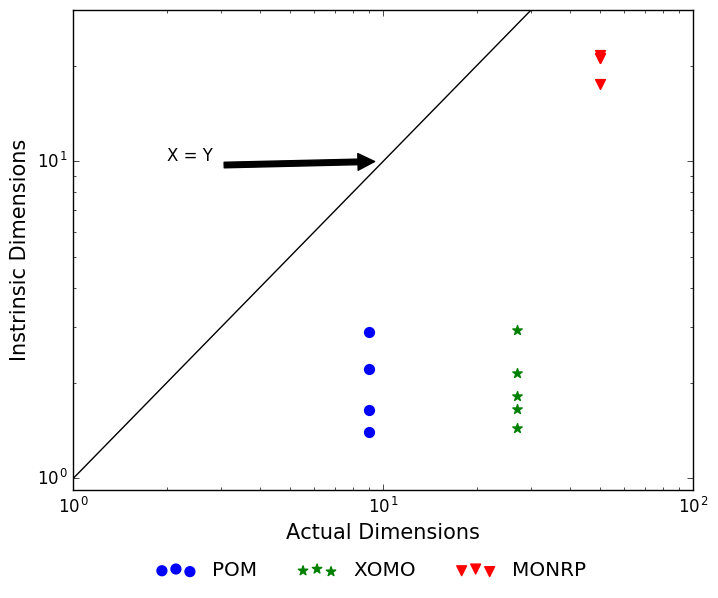}
\caption{The (log of) actual dimensions are shown on the x-axis and (log of) intrinsic dimensionality is shown on the y-axis. 
The intrinsic dimensionality of the systems are much lower than the actual dimensionality (number of columns in the dataset) - this can be easily observed since all the points are lower the line $x=y$, which means points where actual dimensions is equal to intrinsic dimension.}
\label{fig:underlying_d}
\end{figure}

\subsection{Testing technique}
Given our hypothesis that the problem space lies in a lower dimensional hyperplane ---  it is imperative to demonstrate that the intrinsic dimensionality of the configuration space is less than the actual dimension. To formalize this notion, we borrow the concept of correlation dimension from the domain of physics~\cite{grassberger2004measuring}. The correlation dimension of a dataset with $k$ items is found by computing the number of items found at distance within radius $r$ (where r is the Euclidean distance between two configurations) while varying $r$. This is then normalized by the number of connections between $k$ items to find the expected number of neighbors at distance $r$. This can be written as:
\begin{equation}
    C(r) = \frac{2}{k(k-1)} \displaystyle\sum_{i=1}^{n} \displaystyle\sum_{j=i+1}^{n} I(||x_i, x_j|| < r) \\  
 \end{equation} 
$$
where:
    I(x < y) = \begin{cases}
        1,  \text{ if x \textless y}\\
        0,  \text{ otherwise}\\
    \end{cases}
$$

Given the dataset with $k$ items and range of distances [$r_0$--$r_{max}$], we estimate the intrinsic dimensionality as the mean slope between $\ln(C(r))$ and $\ln(r)$.

\subsection{Evaluation}

On the space of problems explored in this paper, we observe that {the intrinsic dimensionality of the problems is much lower than the actual dimension}. Figure \ref{fig:underlying_d} presents the intrinsic dimensionality along with the actual dimensions of the software systems. 
If we look at the intrinsic dimensionality and compare it with the actual dimensionality, then it becomes apparent that the configuration space lies on a lower dimensional hyperplane. For example, the MONRP family of problems has 50-dimensional decision space, but the average intrinsic dimensionality of the space is just $20.39$ 
(this is a fractal dimension). At the heart of  \sway{} is WHERE (a spectral clusterer), 
which uses the approximation of the first principal component to divide the configuration space and hence can take advantage of the low intrinsic dimensionality. 

As a summary, our observations indicate that the intrinsic dimension of the problem space explored in this paper is much lower than its actual dimension. 

Hence, recursive top-down clustering based on the intrinsic dimensions rather than the actual dimension is be more effective. 
In other words, decision space with similar performance values lie closer to the intrinsic hyperplane, compared to the actual dimensions.

\label{rvt:3a1}\rvt{3a1}With this hypothesis, \sway{} recursively cluster candidates basing on intrinsic dimensions only. Also, with the assumption genotype - phenotype space mapping assumption
(see \S\ref{sect:swayo}), \sway{} does not need to evaluate every candidates, making \sway{} so quickly to terminate.

\section{Threats to Validity}\label{sect:threats}
\subsection{Optimizer bias}

There are theoretical reasons to conclude that it is impossible to
show that any one optimizer {\em always} performs best. Wolpert and
Macready~\cite{wolpert97} showed in 1997 that no optimizers
necessarily work better than any other for all possible optimization
problems\footnote{``The computational cost of finding a solution,
  averaged over all problems in the class, is the same for any
  solution method.  No algorithm therefore offers a short
  cut.''~\cite{wolpert97}}.

In this study, we compared {\sc Sway} framework with NSGA-II, SPEA2, GALE for
XOMO, POM3 and MONRP case studies. We selected those learners
since:
\bi
\item
  The literature review of Sayyad et al.~\cite{sayyad13b}
  reported that NSGA-II and SPEA2 were widely used in the SBSE literature;
\item
  GALE is the successor of  \sway{} and this study was partly to show how sampling feature of GALE is the reason why GALE is competitive against widely used evolutionary algorithms.
  \ei
  
That said, there exist many other optimizers. For examples,
NSGA-III is an improved version of NSGA-II, which can get better
diversity of the results. Such kind of optimizers might perform better
than {\sc Sway} method.
Also, for some specific problem, researchers might propose some
modified version of MOEAs. For such kind of modified/optimized algorithms, {\sc Sway} might
not perform as well as them.

\subsection{Sampling bias}

We used the XOMO, POM3 and MONRP SE problems
for our case studies. We found that in all of these problems, \sway{}
can perform as well (or better) as other MOEAs. However, there are many other
optimization problems in the area of SE.

It is very difficult to find the representatives sample of models
which covers all kinds of models. Some problems might have other
properties which make it different from any problem we tested in this
study. For example, project planning problem is an open question up to
now. In the project planning problem, there are constraints which
organized as the directed acyclic graph(DAG). None of problems we
tested with such kind of constraints.

For this issue of sampling bias (and for optimizer bias),
we cannot  explore \textit{all} problems
here. What we can do:
\bi
\item
  Detail our current work and encourage other researchers
  to test our software on a wider range of problems;
\item
In future work, apply {\sc Sway} when we come across a new problem
and compare them to the existed algorithms.
\ei

\subsection{Evaluation bias}

We evaluated the results through spread and hypervolume and number
of evaluations. 
There are many other measures which are adopted in the community of
software engineering. For example, a widely used evaluation measures
for the multi-objective optimization problem is the inverted
generational distance (\textit{igd})
indicator~\cite{van1998multiobjective}.

Using various measures might lead to different conclusions. This
threatens our conclusion. A comprehensive analysis using other
measures is left to the future work.

\section{Future Work}\label{sect:future}
\rv{
We focused on three SBSE problems -- two for software management (XOMO and POM), and another for requirement engineering (MONRP). In the future, we will explore problems in software testing--  an important area in search
based software engineering. 
\label{rvt:2c}\rvt{2c} Recently, Panichella et al.~\cite{panichella2017automated, panichella2015reformulating} addressed that software testing should be extended into a multi-objective optimization, instead of just consideration the ``branch coverage rate'', making such problems trickier.

Another area for future work is to explore
sampling methods for single-objective  problems.
In this paper, we were focusing on multi-objective problems. For multi-objective problems, one essential target is the diversity of results; while in single-objective optimization problems, 
we hope to find out single {\it best}  configurations. Perhaps
in that simpler domain, methods like \sway{} outperform existing state-of-the-art methods.
}

\label{rvt:3a2}\rvt{3a2} One assumption of \sway{} is the genotype-phenotype mapping of 
the explored problem space (see \S\ref{sect:swayo}). In the future, we could possibly improve  \sway{} via regressions between specific 
variables (instead of whole decision variables) and objectives.

\label{rvt:3b}\rvt{3b}Finally, in this paper, \sway{} selects from a large pool of candidates generated initially. An alternate strategy
that might be useful would be to build   sample candidates adaptive
(e.g. using Bayesian parameter optimization or Estimation of Distribution Algorithms~\cite{larranaga2002review}). Such an approach might help  avoid  unnecessary candidate generation.

\section{Conclusion}
In this paper we introduce an alternative to evolutionary algorithms for solving SBSE problems. 
The major drawback of EAs is the slow convergence of solutions, which means it requires too many evaluations/measurements  to solve a problem. To overcome this limitation we introduce  \sway{}, which explores the concept of sampling. 
The key idea of \sway{} is to use the underlying dimension of the problem  to recursively sample the solutions, while only evaluating  the extreme points in the decision space. This strategy helps us to reduce the number of evaluations to find ``interesting'' solutions. 

We evaluated \sway{} on three SE  models, which includes models with continuous as well as discrete  space. Our approach achieves similar performance (hypervolume and spread), if not better, when compared to other sophisticated optimizers using far fewer evaluations ($\frac{1}{20}$ of evaluations used by traditional EAs). 

We demonstrate how \sway{} can be adapted to suit various problems. As an example, we show how barebone \sway{}, which is applicable for problems on the continuous domains, can be adapted for constrained problem in discrete space. Furthermore, we explore the possibility of using the results from \sway{} to ``super-charge'' traditional EAs. We find that supercharging traditional EAs with solutions from \sway{} does not improve their performance significantly.

Hence, our observations lead us to believe that sampling is a cheaper alternative to the more expensive and sophisticated evolutionary techniques. We also believe that techniques such as \sway{}, which exploit the underlying dimension of the problem space, would be very useful to solve problems, where the practitioners are not at the liberty to perform thousands of evaluations.  Observation is in line with a current trend in   machine learning research. 
Dasgupta
and Freund~\cite{Dasgupta2008} comment that:
\begin{quote}
A recent positive  in
that field  has been the realization that a lot of data
which superficially lie in a very high-dimensional space $R^D$,
actually have low intrinsic dimension, in the sense of lying
close to a manifold of dimension $d \ll D$.
\end{quote}

With this paper we want to propagate the message of ``easy over hard'' and we believe \sway{} is a step towards this direction.
\rv{Even in domains where \sway{} does not
outperform  other  evolutionary algorithms, we would still suggest
researchers to first apply \sway{}, since this is a simple and fast algorithm which can
rapidly produce baseline results.
}

\balance
\bibliographystyle{plain}
\bibliography{main}

\end{document}